\documentclass[preprint]{aastex}
\usepackage{amsmath}
\usepackage{color}
\usepackage{ulem}
\begin{document}

\title{No Keplerian Disk $>$10 AU around the Protostar B335: Magnetic Braking or Young Age?}

\author{Hsi-Wei Yen\altaffilmark{1}, Shigehisa Takakuwa\altaffilmark{1}, Patrick M. Koch\altaffilmark{1}, Yusuke Aso\altaffilmark{2}, Shin Koyamatsu\altaffilmark{2,3}, Ruben Krasnopolsky\altaffilmark{1}, and Nagayoshi Ohashi\altaffilmark{1,3}}

\altaffiltext{1}{Academia Sinica Institute of Astronomy and Astrophysics, P.O. Box 23-141, Taipei 10617, Taiwan; hwyen@asiaa.sinica.edu.tw} 
\altaffiltext{2}{Department of Astronomy, Graduate School of Science, The University of Tokyo, 7-3-1 Hongo, Bunkyo-ku, Tokyo 113-0033, Japan}
\altaffiltext{3}{Subaru Telescope, National Astronomical Observatory of Japan, 650 North A'ohoku Place, Hilo, HI 96720, USA}

\begin{abstract}
We have conducted ALMA cycle 2 observations in the 1.3 mm continuum and in the C$^{18}$O (2--1) and SO (5$_6$--4$_5$) lines at a resolution of $\sim$0\farcs3 toward the Class 0 protostar B335. 
The 1.3 mm continuum, C$^{18}$O, and SO emission all show central compact components with sizes of $\sim$40--180 AU within more extended components. 
The C$^{18}$O component shows signs of infalling and rotational motion.
By fitting simple kinematic models to the C$^{18}$O data, 
the protostellar mass is estimated to be 0.05 $M_\sun$. The specific angular momentum, on a 100 AU scale, is (4.3$\pm$0.5) $\times$ 10$^{-5}$ km s$^{-1}$ pc. 
A similar specific angular momentum, (3--5) $\times$ 10$^{-5}$ km s$^{-1}$ pc, is measured on a 10 AU scale from the velocity gradient observed in the central SO component, 
and there is no clear sign of an infalling motion in the SO emission. 
By comparing the infalling and rotational motion, 
our ALMA results suggest that the observed rotational motion has not yet reached Keplerian velocity neither on a 100 AU nor even on a 10 AU scale.
Consequently, the radius of the Keplerian disk in B335 (if present) is expected to be 1--3 AU. 
The expected disk radius in B335 is one to two orders of magnitude smaller than those of observed Keplerian disks around other Class 0 protostars. 
Based on the observed infalling and rotational motion from 0.1 pc to inner 100 AU scales, 
there are two possible scenarios to explain the presence of such a small Keplerian disk in B335: magnetic braking and young age.
If our finding is the consequence of magnetic braking, 
$\sim$50\% of the angular momentum of the infalling material within a 1000 AU scale might have been removed, 
and the magnetic field strength on a 1000 AU scale is estimated to be $\sim$200 $\mu$G.
If it is young age, 
the infalling radius in B335 is estimated to be $\sim$2700 AU, corresponding to a collapsing time scale of $\sim$5 $\times$ 10$^4$ yr.

\end{abstract}

\section{Introduction}
Keplerian disks are often observed around young stellar objects and are considered to be sites of planet formation (e.g., Williams \& Cieza 2011). 
Such disks are also observed around sources at even earlier evolutionary stages, Class 0 and I protostars, which are still embedded in dense envelopes (e.g., Tobin et al.~2012b; Takakuwa et al.~2012; Brinch \& J{\o}rgensen 2013; Murillo et al.~2013; Harsono et al.~2014; Yen et al.~2014; Lee et al.~2014; Codella et al.~2014; Chou et al.~2014; Ohashi et al.~2014).
The Keplerian disks around these Class 0 and I protostars have radii ranging from 50 AU to 350 AU and masses from 10$^{-3}$ $M_\odot$ to 10$^{-1}$ $M_\odot$, which are similar to disks around young stellar objects (e.g., Simon et al.~2000; Pi\'etu et al.~2007). 
The comparable disk masses and sizes around Class 0 and I protostars and young stellar objects suggest that Keplerian disks can be well developed at an early evolutionary stage of star formation. 
In our study of gas motions on a 1000 AU scale around 17 Class 0 and 0/I protostars using the Submillimeter Array (SMA), 
we found that 13 out of the 17 sources could exhibit Keplerian disks with a size scale $\gtrsim$100 AU (Yen et al.~2015).
Opposite to this, 
the following group of Class 0 protostars is showing no sign that such 100 AU Keplerian disks have formed: B335 (Yen et al.~2010, 2015), NGC 1333 IRAS 4B (Yen et al. 2013, 2015), NGC 1333 IRAS 2A (Brinch et al.~2009; Maret et al.~2014) and L1157-mm (Yen et al.~2015). 
The protostellar envelopes on a 1000 AU scale around these protostars do not show clear rotational motions, 
suggesting that the specific angular momenta in their envelopes are likely small. 
Because of these small specific angular momenta ($\lesssim$10$^{-4}$ km s$^{-1}$ pc; Yen et al.~2015), 
the outer radii of their Keplerian disks (if present) are likely less than 10 AU. 
Therefore, based on the current studies with limited sample sizes (e.g., Yen et al.~2015), 
one fifth of Class 0 protostars could be without well developed Keplerian disks. 

A Keplerian disk is expected to form when collapsing material rotates fast enough around a protostar to become rotationally supported (e.g., Shu et al.~1987). 
If the angular momentum of the collapsing material is conserved, 
this material will rotate faster as it approaches the protostar.  In this way,  a Keplerian disk is naturally formed (e.g., Ulrich 1976; Cassen \& Moosman 1981; Terebey et al.~1984; Basu 1998; Bate 1998).  
In this scenario, 
the size of a Keplerian disk grows as more angular momentum is transferred to the disk by the collapsing material.
A Keplerian disk with a radius of 100 AU can then form within 10$^5$ yr (e.g., Yen et al.~2015). 
This is comparable to the time scale of the Class 0 stage (Enoch et al.~2009; Dunham et al.~2014).
On the contrary,
if magnetic fields are incorporated,  
theoretical studies show that the magnetic fields can slow down the infalling velocity of collapsing material, remove its angular momentum, and suppress the formation and growth of Keplerian disks. 
This mechanism can be very efficient. As a result, no Keplerian disk with a size larger than 10 AU can form (e.g., Allen et al.~2003; Mellon \& Li 2008, 2009; Machida et al.~2011; Li et al.~2011; Dapp et al.~2012). 
Recent high-resolution magnetohydrodynamics (MHD) simulations incorporating ambipolar diffusion and ohmic dissipation show that a small rotationally-supported disk with a size of a few AU can form at the early evolutionary stage of star formation (e.g., Machida et al.~2014; Tomida et al.~2015). 
Although these simulations do not continue to follow the growth of that small disk due to computational limitations, 
these studies expect that the disk size will eventually grow to tens of AU with a proceeding collapse.
However, Keplerian disks with sizes larger than tens of AU have been observed around several Class 0 and I protostars. 
Observationally, it is, thus, not yet clear when and whether or not efficient magnetic braking is occurring during the star formation process.

To investigate the formation process of Keplerian disks around Class 0 protostars, 
we conducted observations with the Atacama Large Millimeter/Submillimeter Array (ALMA) toward B335, which is one of the Class 0 sources showing no clear sign of rotational motion on a 1000 AU scale (Yen et al.~2010, 2015). 
B335 is an isolated Bok globule associated with an infrared source (IRAS 19347+0727) having a bolometric luminosity of 1.5 $L_{\sun}$ at a distance$\footnotemark$ of 150 pc (Keene et al.~1980, 1983; Stutz et al.~2008). 
The CO outflow associated with B335 is extended along the east--west direction and has an opening angle of $\sim$45$\degr$--60$\degr$ (e.g., Hirano et al.~1988; Yen et al.~2010; Hull et al.~2014). 
The inclination angle of the outflow axis is estimated to be $\sim$10$\degr$ from the plane of the sky, suggesting that B335 is almost an edge-on source (Hirano et al.~1988).
Single-dish observations of B335 show signatures of infalling motions on a scale of thousands of AU (Zhou et al.~1993, 1995; Choi et al.~1995; Evans et al.~2005) and rotational motions on a 0.1 pc scale (Saito et al.~1999; Yen et al.~2011; Kurono et al.~2013). 
Interferometric observations of the inner envelope on a 1000 AU scale also show the presence of infalling motions (Chandler \& Sargent 1993; Saito et al.~1999; Yen et al.~2010; Kurono et al.~2013) but no clear sign of rotational motions (Yen et al.~2010, 2015). 
The absence of a clear rotational motion on a 1000 AU scale suggests that the outer radius of the Keplerian disk (if present) is likely $<$10 AU (Yen et al.~2010).  
Such a small Keplerian disk further suggests that magnetic braking might be efficient in B335 or that B335 is at an early stage of disk formation (Yen et al.~2011, 2013, 2015).
Therefore, B335 is a good target to study how collapsing material carries angular momenta to the vicinity of a protostar and forms a Keplerian disk. 

\footnotetext{The most recent measurement of the distance to B335 is $\sim$90--120 pc by Olofsson \& Olofsson (2009), which is comparable to the distance of 150 pc measured by Stutz et al.~(2008). In the present paper, a distance of 150 pc is adopted for comparison with previous studies.}

In the present paper, we report our observational results of the 1.3 mm continuum, C$^{18}$O (2--1; 219.560358 GHz), and SO (5$_6$--4$_5$; 219.949433 GHz) emission in B335 obtained with ALMA. 
C$^{18}$O can be abundant in protostellar envelopes as the temperature can increase, due to heating by protostars and accretion processes, to above its evaporation      
temperature of $\sim$20 K. 
Besides, the C$^{18}$O emission is typically optically thin ($\tau \lesssim 0.3$) in protostellar sources as inferred from the observed C$^{18}$O (3--2) and (2--1) line ratios ($\sim$1--2; e.g., Hogerheijde et al.~1998). 
Hence, the C$^{18}$O line can be used to trace the kinematics of protostellar envelopes. 
Compact SO emission on a scale of a few hundred AU around protostars is thought to be related to accretion shocks where collapsing material falls onto Keplerian disks, as revealed by recent ALMA observations (e.g., Sakai et al.~2014; Yen et al.~2014; Ohashi et al.~2014). 
Therefore, the SO line can potentially be used to study the transitional region from infalling to dominating rotational motion. 
In this paper, we investigate the gas kinematics in the innermost regions within a 100 AU scale around the protostar in B335 using the C$^{18}$O (2--1) and SO (5$_6$--4$_5$) lines with our high-resolution ALMA observations ($\sim$0\farcs2--0\farcs3 or $\sim$30--45 AU).
Previous single-dish and interferometric observations revealed the gas kinematics on scales from 10,000 AU to 1000 AU in B335.  
With these previous observational results and our new ALMA results, 
we study the infalling motion and the angular momentum transfer from the scale of the dense core to the vicinity around the protostar. 
Our results are discussed in comparison with theoretical models with and without magnetic fields to investigate disk formation.

\section{Observations}
The ALMA Band 6 observations of B335 were made with 34 antennas during the cycle-2 observing period on September 2, 2014.
The pointing center was $\alpha$(J2000) = $19^{h}37^{m}00\fs89$, $\delta$(J2000) = $7\arcdeg34\arcmin10\farcs0$.  
The on-source integration time on B335 was $\sim$25 minutes. 
The projected $uv$ distances range from $\sim$25 k$\lambda$ to $\sim$800 k$\lambda$.
Our observations are insensitive to structures larger than $\sim$3\farcs5 ($\sim$525 AU) at a 50\% level (Wilner \& Welch 1994). 
The correlator was configured in the Frequency Division Mode.
Two spectral windows, each with a bandwidth of 234 MHz, were assigned to the C$^{18}$O (2--1) and SO (5$_6$--4$_5$) lines with 3840 spectral channels in each window, resulting in a channel width of 61 kHz. 
Two spectral windows with of a total bandwidth of 4 GHz were assigned to the 1.3 mm continuum. 
Calibration of the raw visibility data was performed with the standard reduction script for the cycle-2 data, which uses tasks in Common Astronomy Software Applications (CASA), and without self calibration.
J1751+096 was observed as a bandpass and flux calibrator.
The absolute flux of J1751+096 was measured to be 2.53 Jy at a frequency of 233 GHz in other observations. 
The fluxes at our observing frequencies were extrapolated from this flux.
J1955+1358 was observed as a gain calibrator. 
The calibrated visibility data of the C$^{18}$O and SO lines were Fourier-transformed with the Briggs robust parameter of +0.5 and CLEANed with the CASA task ``clean'' at a velocity resolution of 0.17 km s$^{-1}$. 
For the 1.3 mm continuum, the calibrated visibility data were Fourier-transformed with three different weightings (natural, uniform, and long baseline only  with $uv> 500$ k$\lambda$) and CLEANed. 
The synthesized beams and the rms noise levels of our images are listed in Table \ref{ob}.

\section{Results}
The systemic velocity of B335 was previously estimated to be $V_{\rm LSR} = 8.34$ km s$^{-1}$ from the single-dish spectrum of the C$^{18}$O (2--1) emission at an angular resolution of 33$\arcsec$ (Yen et al.~2011). 
In the present paper, we adopt this systemic velocity. 
Velocities relative to the systemic velocity are labeled as $\Delta V$. 
Velocity channel maps are shown in Appendix \ref{chan} for our observed emission lines. 

\subsection{1.3 mm Continuum Emission}
\begin{figure}
\epsscale{1}
\plotone{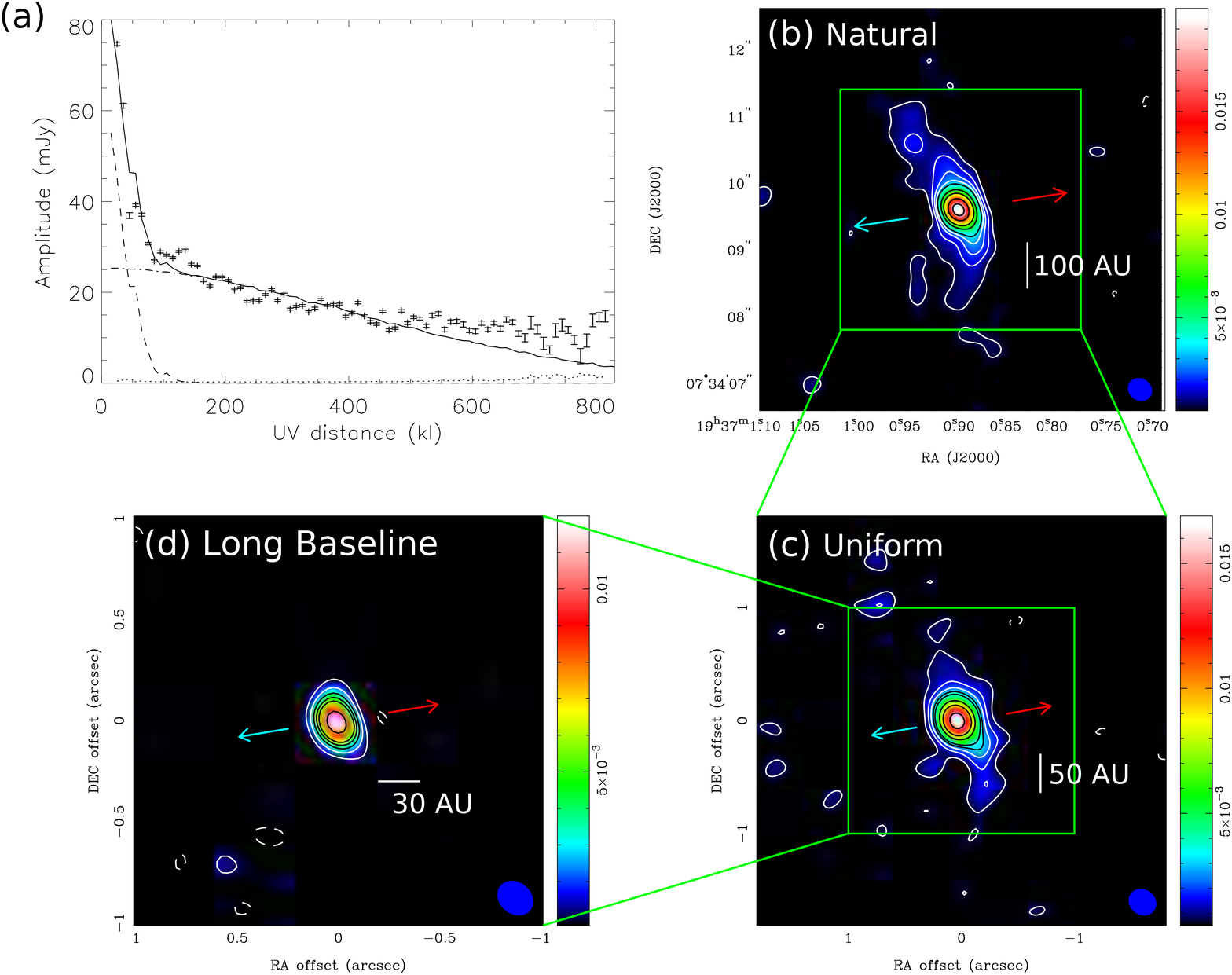}
\caption{(a) Visibility amplitude as a function of $uv$ distance of the 1.3 mm continuum emission in B335 observed with ALMA. Dashed and dot-dashed lines present the visibility amplitude profiles of the fitted extended and central compact Gaussian components, respectively. The combined visibility amplitude profile of these two components is shown as a solid line. A dotted line shows the expected amplitude for a zero signal. (b)--(d) 1.3 mm continuum images of B335 generated with natural weighting, uniform weighting, and only long baselines with $uv$ distance larger than 500 k$\lambda$. Green boxes show areas of images with subsequently higher resolutions. Filled ellipses at the bottom-right corner in each panel present the beam sizes. Red and blue arrows denote the directions of the redshifted and blueshifted outflows, respectively. Contour levels are 3$\sigma$, 6$\sigma$, 9$\sigma$, 12$\sigma$, 15$\sigma$, 25$\sigma$, 35$\sigma$, and then in steps of 20$\sigma$. The beam sizes and the 1$\sigma$ noise levels are summarized in Table \ref{ob}.}
\label{con}
\end{figure}

Figure \ref{con}a presents visibility amplitudes as a function of $uv$ distances for the 1.3 mm continuum emission in B335 observed with ALMA.
The visibility amplitude rapidly decreases as the $uv$ distance increases from $\sim$30 k$\lambda$ to $\sim$100 k$\lambda$.
The profile then becomes shallower for $uv$ distances beyond $\sim$100 k$\lambda$. 
This visibility amplitude function suggests the presence of two components in the 1.3 mm continuum emission: an extended and a compact one. 
Consequently, we fit two Gaussians to the 1.3 mm continuum in the $uv$ plane to derive the basic parameters of these two components. 
The compact component is centered at $\alpha$(J2000) = 19$^{h}$37$^{m}$0$\fs$89, $\delta$(J2000) = 7\arcdeg34\arcmin09$\farcs$6. 
In the present paper, this position is adopted as the protostellar source location.
Its total flux, de-convolved full width half maximum (FWHM), and position angle are 25.3$\pm$0.3 mJy, 0\farcs24$\pm$0\farcs01 $\times$ 0\farcs13$\pm$0\farcs01 (36$\pm$2 AU $\times$ 20$\pm$2 AU), and 16\degr$\pm$2\degr, respectively. 
This compact component is elongated perpendicularly to the outflow axis which has a position angle of $\sim$99$\degr$ (e.g., Hull et al.~2014).
Hence, it likely traces the flattened structure around the protostar. 
The extended component is centered 0\farcs25 east to the center of the compact component. 
Its total flux, de-convolved FWHM, and position angle are 62.2$\pm$2.8 mJy, 3\farcs34$\pm$0\farcs13 $\times$ 1\farcs41$\pm$0\farcs06 (501$\pm$20 AU $\times$ 212$\pm$9 AU), and 25\degr$\pm$1\degr, respectively. 
A comparison between the observed visibility amplitudes and the two Gaussian components is shown in Figure \ref{con}a. 
The observed visibility amplitude function is flatter than the Gaussian of the extended component at the longer $uv$ distances ($\sim$500--800 k$\lambda$), 
showing an excess in amplitude at $uv$ distances beyond $\sim$500 k$\lambda$. 
This observed flatter profile suggests that the inner envelope exhibits even more compact structures which cannot be described by a simple Gaussian.

Figure \ref{con} b--d present ALMA 1.3 mm continuum images of B335, made with natural and uniform weighting, and with baselines longer than 500 k$\lambda$ only. 
The natural-weighting image shows a central compact component with an apparent size of $\sim$0\farcs5 ($\sim$75 AU) within an elongated structure with a length of $\sim$1\farcs5 ($\sim$225 AU). 
The elongated structure extends along the northeast--southwest direction with a position angle of 20\degr--30\degr.
Its most southern part is bent and points toward the southeast (Fig.~\ref{con}b). 
In the image with uniform weighting, 
the central compact component is more clearly resolved (Fig.~\ref{con}c), 
and shows an elongation perpendicular to the outflow.
This component corresponds to the compact component identified in the visibility amplitude function in Figure \ref{con}a.
Figure \ref{con}d presents a 1.3 mm continuum image generated using only the data with baselines longer than 500 k$\lambda$ and uniform weighting. 
This image demonstrates that the 1.3 mm continuum emission is unresolved and appears to be a point source even with baselines up to $\sim$800 k$\lambda$ (corresponding to $\sim$0\farcs3 or 45 AU), suggesting the presence of even more compact structures in the central component.
The flux of this unresolved component is estimated to be $\sim$13.2$\pm$0.6 mJy.

The mass of the circumstellar material traced by the 1.3 mm continuum emission ($\tbond$$M_{\rm 1.3mm}$) can be estimated as
\begin{equation}
M_{\rm 1.3mm} = \frac{F_{\rm 1.3mm}d^{2}} {\kappa_{\rm 1.3 mm} B(T_{\rm dust})},
\end{equation}
where $F_{\rm 1.3mm}$ is the total 1.3 mm flux, $d$ is the distance to the source, $\kappa_{\rm 1.3 mm}$ is the dust mass opacity at 1.3 mm, $T_{\rm dust}$ is the dust temperature, and $B(T_{\rm dust})$ is the Planck function at a temperature of $T_{\rm dust}$. 
Under the assumption that the wavelength ($\tbond$$\lambda$) dependence of the dust mass opacity ($\equiv \kappa_{\lambda}$) is $\kappa_{\lambda} = 0.1 \times (0.3\ {\rm mm}/\lambda)^{\beta}$ cm$^{2}$ g$^{-1}$ (Beckwith et al.~1990), the mass opacity at 1.3 mm is 0.014 cm$^{2}$ g$^{-1}$ with $\beta = 1.3$ (Chandler \& Sargent 1993) and a gas-to-dust mass ratio of 100. 
$T_{\rm dust}$ on a scale of $\sim$600 AU in B335 has been estimated to be $\sim$30 K (Chandler \& Sargent 1993). 
Therefore, $M_{\rm 1.3mm}$ of the extended component with a size of $\sim$500 AU identified in the visibility amplitude function is estimated with $T_{\rm dust}$ of 30 K to be $\sim$0.012 $M_\odot$.
The compact component of the 1.3 mm continuum emission likely traces the innermost region on a scale of $\sim$40 AU.
Hence, $T_{\rm dust}$ in the inner region is expected to be higher. 
Assuming that $T_{\rm dust}$ increases as a power-law function with a power-law index of $-0.4$ as the radius decreases (e.g., Shirley et al.~2000), 
$T_{\rm dust}$ on a scale of $\sim$40 AU is estimated to be $\sim$90 K. 
The radiative transfer models of infalling envelopes also show that $T_{\rm dust}$ can be as high as 100 K within a radius of $\sim$25 AU (Visser et al.~2009). 
In addition to the increase in the temperature along the radial direction, 
the inner flattened envelope can also exhibit a vertical temperature gradient,
where the temperature is lower in the mid-plane and higher in the upper layers (e.g., Visser et al.~2009).
Therefore, here $M_{\rm 1.3mm}$ of the compact and unresolved components is estimated with $T_{\rm dust}$ of 30--100 K. 
We note that with the same $F_{\rm 1.3mm}$,  $T_{\rm dust}$ of 100 K gives a factor of $\sim$3.8 smaller $M_{\rm 1.3mm}$ as compared to $T_{\rm dust}$ of 30 K.
As a result, $M_{\rm 1.3mm}$ of the compact component is $\sim$(1.3--5.4) $\times$ 10$^{-3}$ $M_\sun$
and that of the unresolved component is $\sim$(0.7--2.6) $\times$ 10$^{-3}$ $M_\sun$. 
The results of the 1.3 mm continuum emission are summarized in Table \ref{dust}.
 
\subsection{C$^{18}$O (2--1) Emission}
\begin{figure}
\epsscale{1}
\plotone{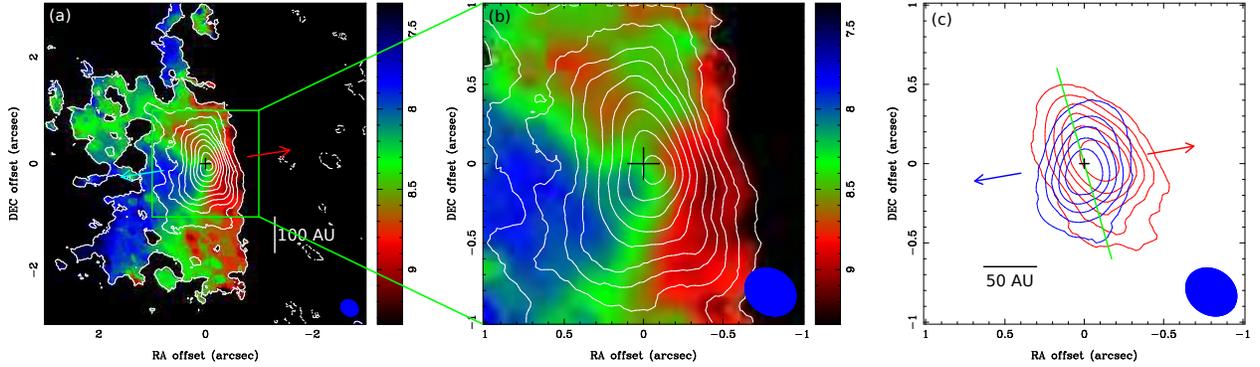}
\caption{(a) Moment 0 map (contours) overlaid on the moment 1 map (color scale, units in km s$^{-1}$) of the C$^{18}$O (2--1) emission in B335. The green box shows the area of panel (b) and (c). (b) Same as (a) but zooming in on the central 2$\arcsec$ region. (c) Moment 0 map of the high-velocity redshifted and blueshifted C$^{18}$O emission in B335 with $\Delta V > 1.2$ km s$^{-1}$. Filled ellipses at the bottom-right corners present the beam sizes. Crosses indicate the protostellar positions. Red and blue arrows denote the directions of the red- and blueshifted outflows, respectively. The green line shows the direction of the major axis of the 1.3 mm continuum emission. Contour levels are 3$\sigma$, 6$\sigma$, 9$\sigma$, 12$\sigma$, 15$\sigma$, and then in steps of 5$\sigma$. 1$\sigma$ noise levels in panel (a) and (c) are 3.4 and 1.7 mJy beam$^{-1}$, respectively.}
\label{c18o}
\end{figure}

Figure \ref{c18o}a presents the intensity-integrated (moment 0) map overlaid on the intensity-weighted mean-velocity (moment 1) map of the C$^{18}$O (2--1) emission in B335. 
The C$^{18}$O emission shows a central component within a more diffuse component which extends toward the east. 
The central component is elongated along the north--south direction and has an apparent size of $\sim$2$\arcsec$ ($\sim$300 AU).
A two-dimensional Gaussian is fitted to the central compact component to estimate its size. 
Pixels below 10$\sigma$ are excluded in the fitting to avoid a contamination from the diffuse component. 
The de-convolved FWHM of the central component is derived to be 1\farcs2$\pm$0\farcs1 $\times$ 0\farcs75$\pm$0\farcs05 (180$\pm$15 AU $\times$ 113$\pm$8 AU) with a position angle of $\sim$4\degr$\pm$5\degr, 
which is perpendicular to the outflow axis.  
The derived peak position is consistent with the protostellar position within the uncertainty. 

The C$^{18}$O (2--1) emission shows an overall velocity gradient along the east--west direction, 
identical to the one observed with the SMA at a lower resolution of $\sim$4$\arcsec$ (Yen et al.~2010). 
The diffuse emission is blueshifted and located to the east, 
similar to the blueshifted outflow in B335. 
The eastern diffuse component is likely associated with the outflow activity.  
The central component, which is elongated perpendicularly to the outflow, likely traces the flattened envelope around the protostar, similar to the 1.3 mm continuum emission.
Figure \ref{c18o}b shows the central region of the C$^{18}$O emission.
The central component also exhibits a velocity gradient along the east--west direction. 
This velocity gradient along the outflow axis in the flattened envelope can be due to the infalling motion or the contamination from the outflow, as seen in other protostellar sources (e.g., Ohashi et al.~1997; Momose et al.~1998; Arce \& Sargent 2006). 
Besides, the central component likely also exhibits a velocity gradient along the north--south direction. 
In this central component, the C$^{18}$O (2--1) emission is more redshifted to the northeast and is more blueshifted to the southeast. 
This velocity gradient perpendicular to the outflow axis in the flattened envelope was not identified in the SMA C$^{18}$O (2--1) observation because of its lower resolution of $\sim$4$\arcsec$ (Yen et al.~2010).  
This central gradient can be a sign of a rotational motion on a 100 AU scale in B335. 
In contrast, 
the outer extended C$^{18}$O (2--1) emission is more redshifted to the south and more blueshifted to the north. 
However, the size of the outer extended component ($\sim$4\arcsec) is larger than the maximum angular scale that our ALMA observations can recover. 
Hence, its full velocity structures are still uncertain.

At the high velocities, $\Delta V > 1.2$ km s$^{-1}$, the C$^{18}$O emission appears to be more compact with a minimum contamination from the more extended component (see Appendix \ref{chan}). It, therefore, likely traces the innermost region.
Figure \ref{c18o}c presents the moment 0 map of the high-velocity C$^{18}$O emission at $\Delta V > 1.2$ km s$^{-1}$. 
The high-velocity redshifted C$^{18}$O emission is located to the northwest, and the high-velocity blueshifted C$^{18}$O emission to the southeast of the protostellar position. 
The separation between 
their peaks is measured to be $\sim$0\farcs11$\pm$0\farcs01 ($\sim$16.5$\pm$1.5 AU), 
and the position angle of the axis passing through their peaks is 128\degr$\pm$10\degr, 
suggesting a velocity gradient with a magnitude of $\sim$0.2 km s$^{-1}$ AU$^{-1}$ along the outflow direction.
In other protostars observed with ALMA in the C$^{18}$O (2--1) line at sub-arcsecond resolutions, 
such as VLA 1623 A (Murillo et al.~2013), L1527 IRS (Ohashi et al.~2014), and L1489 IRS (Yen et al.~2014), 
the high-velocity redshifted and blueshifted C$^{18}$O (2--1) components are aligned along the major axes of the 1.3 mm continuum emission and perpendicularly to the outflow axes, 
suggesting that the gas motions in the innermost regions on a 100 AU scale in these sources are dominated by rotational motions. 
This feature is not seen in B335. 
On the contrary, in B335, the high-velocity redshifted and blueshifted C$^{18}$O (2--1) components are aligned along the minor axis of the 1.3 mm continuum emission and the outflow axis.  
Therefore, the innermost envelope on a 100 AU scale in B335 is likely not dominated by rotational motion but more likely infalling with only, if at all, a minimum rotational motion.

\subsection{SO (5$_6$--4$_5$) Emission}
Figure \ref{somap}a presents the moment 0 map overlaid on the moment 1 map of the SO (5$_6$--4$_5$) emission in B335. 
The SO emission shows a central compact component with an apparent size of $\sim$0\farcs5 ($\sim$75 AU) and an extended structure with a length of $\sim$1$\arcsec$ elongated along the northeast--southwest direction. 
These features are similar to those seen in the 1.3 mm continuum emission. 
A two-dimensional Gaussian is fitted to the central compact component. 
Pixels below 20$\sigma$ are excluded in the fitting to avoid the contamination from the extended structure. 
The de-convolved FWHM and position angle of the central compact component are derived to be 0\farcs29$\pm$0\farcs01 $\times$ 0\farcs22$\pm$0\farcs01 (44$\pm$2 AU $\times$ 33$\pm$2 AU) and 32\degr$\pm$6\degr, respectively. 
Its peak position is consistent with the protostellar position within the uncertainty. 
Its orientation is also perpendicular to the outflow, as is the central compact C$^{18}$O component.
The size of the central component of the SO emission is a factor of four smaller than that of the C$^{18}$O emission.
It is detected at even higher velocities up to $\Delta V \sim $ 2.7--3.4 km s$^{-1}$ while the C$^{18}$O emission is detected at $\Delta V \lesssim $ 2.0--2.3 km s$^{-1}$ (Fig.~\ref{c18ochan1s} and \ref{sochan1}).
These results suggest that the central component of the SO emission likely arises from a more inner region than the C$^{18}$O emission.
The central compact SO component is more redshifted to the northeast and more blueshifted to the southwest, suggesting a velocity gradient along the northeast--southwest direction.
That is different from the C$^{18}$O emission having an overall velocity gradient along the east--west direction.
Therefore, the gas motion in the innermost region over 40 AU traced by the SO emission is likely different from that in the outer region over 180 AU traced by the C$^{18}$O emission.

Figure \ref{somap}b presents the moment 0 map of the high-velocity SO emission at $\Delta V > 1.2$ km s$^{-1}$, 
to separate the features of the central compact component from the contamination of the extended component.
Two-dimensional Gaussians are fitted to the redshifted and blueshifted components to measure their peak positions. 
The redshifted velocities show a peak that is located more to the northeast with respect to the peak of the blueshifted velocities. 
Their separation is 
$\sim$0\farcs045$\pm$0\farcs004 ($\sim$6.8$\pm$0.6 AU)
with a position angle of the axis passing through their peaks of $\sim$30\degr$\pm$6\degr.
From the velocity difference of the high-velocity redshifted and blueshifted components ($\sim$4.1 km s$^{-1}$) and their separation, 
the magnitude of this velocity gradient is estimated to be $\sim$0.6 km s$^{-1}$ AU$^{-1}$.
The high-velocity redshifted and blueshifted SO emission is aligned along the major axis of the 1.3 mm continuum emission and perpendicularly to the outflow, 
different from the high-velocity C$^{18}$O emission.  
The presence of high-velocity components within a 100 AU scale aligned perpendicularly to outflows is considered as a signature of dominant rotational motion in the vicinity around protostars, 
as observed in other protostellar sources, such as VLA 1623 A (Murillo et al.~2013) L1527 IRS (Ohashi et al.~2014), and L1489 IRS (Yen et al.~2014). 
However, 
the separation between the high-velocity redshifted and blueshifted components in B335 is an order of magnitude smaller than those in VLA 1623 A, L1527 IRS, and L1489 IRS which show a separation of $\sim$0\farcs5--1\arcsec ($\sim$70--120 AU).
Therefore, the results of the SO emission suggest
that compact rotational motion is likely present on small scales of $\sim$40 AU in B335. 
Moreover, its velocity magnitude is smaller than those in other protostellar sources. 

\section{Gas Kinematics}
\subsection{Infalling and Rotating Envelope Traced by the C$^{18}$O (2--1) Emission}\label{c18omotion}
\subsubsection{Observed C$^{\it 18}$O P--V diagrams}
\begin{figure}
\epsscale{1}
\plotone{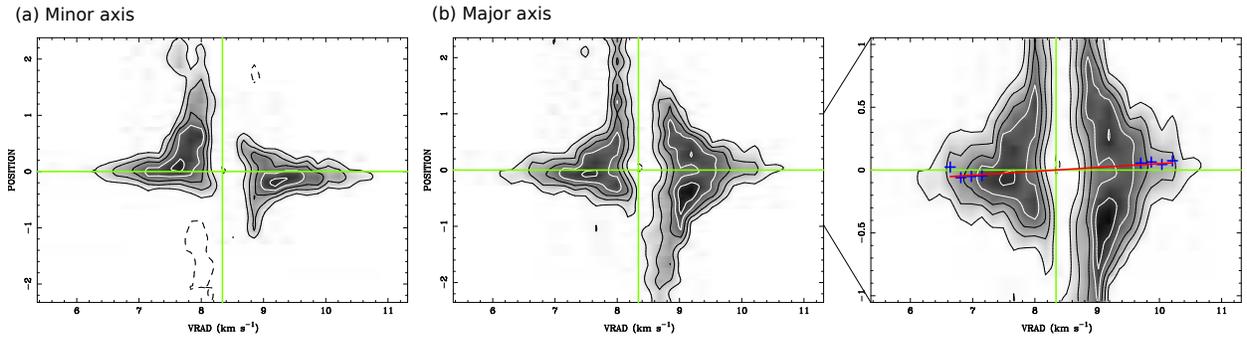}
\caption{P--V diagrams of the C$^{18}$O emission along the minor (left panel) and major (right panels) axes of the 1.3 mm continuum emission. The vertical axis presents position in arcseconds. The horizontal axis is line-of-sight velocity (VRAD) in units of km s$^{-1}$. In the rightmost panel, crosses show the peak positions measured in different velocity channels in the P--V diagram. The red solid line shows the velocity gradient derived from these peak positions at different velocities. Green vertical and horizontal lines denote the systemic velocity and the protostellar position, respectively. Contour levels are from 3$\sigma$ in steps of 3$\sigma$, where 1$\sigma$ is 3.8 mJy beam$^{-1}$. 
}
\label{c18opv}
\end{figure}

Figure \ref{c18opv} presents the position--velocity (P--V) diagrams of the C$^{18}$O emission along the major and minor axes of the 1.3 mm continuum emission to reveal the details of the velocity structures. 
The P--V diagram along the minor axis shows a velocity gradient, where the southeastern region (offset $>$ 0) is more blueshifted and the northwestern region (offset $<$ 0) redshifted (Fig.~\ref{c18opv}a). 
In addition, the low-velocity emission at  $\Delta V < 1$ km s$^{-1}$ is more extended ($\gtrsim$1\arcsec) than the high-velocity emission, 
suggesting that the velocities of the C$^{18}$O component increase as the radii decrease. 
This is different from the velocity structures of the outflow in B335 traced by the $^{12}$CO (2--1) emission in the SMA observations (Yen et al.~2010). 
The low-velocity ($V \lesssim 6$ km s$^{-1}$) $^{12}$CO outflow observed with the SMA exhibits higher velocities at its outer radii, as shown in its P--V diagram along the outflow axis.
Hence, the velocity structures of the C$^{18}$O emission in the inner 1$\arcsec$ region likely have minimum contamination from the outflow, 
while the C$^{18}$O emission at the offset $\gtrsim$ 1$\arcsec$ to the east shows a different velocity structure, 
which is possibly due to the outflow contamination. 
In the P--V diagram along the major axis (Fig.~\ref{c18opv}b), 
both the redshifted and blueshifted C$^{18}$O components are distributed in both the northeastern (offset $>$ 0) and southwestern (offset $<$ 0) regions
and show a triangle-shaped structure. 
These features are also observed with ALMA in the C$^{18}$O (2--1) line in L1527 IRS (Ohashi et al.~2014) and in the HCO$^+$ (4--3) line in HH 212 (Lee et al.~2014)
and can be explained with models of infalling flattened envelopes. 
Furthermore, 
the presence of the relative positional offset between redshifted and blueshifted components in the P--V diagram along the major axis of an infalling flattened envelope is a signature that the envelope is also rotating (e.g., Ohashi et al.~1997, Yen et al.~2013). 
This signature is clearly seen in the P--V diagram of the C$^{18}$O (2--1) emission in L1527 IRS and those of the HCO$^+$ (4--3) and C$^{17}$O (3--2) emission in HH 212.
These two Class 0 protostars are found to exhibit Keplerian disks with sizes of tens of AU (Ohashi et al.~2014; Lee et al.~2014; Codella et al.~2014).
In contrast to these sources, there is no evident positional offset between the redshifted and blueshifted components in the P--V diagram along the major axis in B335, 
suggesting that its rotational motion is likely slow. 

Although there is no evident velocity gradient in the P--V diagram along the major axis, 
the high-velocity C$^{18}$O emission indeed shows a velocity gradient along the major axis as seen in Figure \ref{c18o}. This is likely a sign of a rotational motion. 
To examine this velocity gradient, 
we derive the peak position in each velocity channel at $\Delta V > 1.2$ km s$^{-1}$ in the P--V diagram along the major axis, 
shown as blue data points in Figure \ref{c18opv}b. 
The peak positions are derived by fitting Gaussians to the intensity profiles in these channels using the IDL routine {\it gaussfit}.
Only channels with emission peaks above 5$\sigma$ are included. 
Then, a linear relation $V_{\rm LSR} \propto r$ is fitted to these peak positions. 
The minimization of the fitting is done by using the IDL routine {\it MPFIT} (Markwardt 2009).
We repeat this fitting process for a 1000 times, where each time the position angle and the center of the P--V diagram, and the systemic velocity are randomly changed within their 1$\sigma$ uncertainties, which are 2$\arcdeg$, 0\farcs01, and 0.07 km s$^{-1}$, respectively. 
With this we obtain a probability distribution of the magnitude of the velocity gradient. The 1000 iterations are sufficient to stably converge. 
The moment 1 and 2 of the probability distribution are adopted as the estimated magnitude and its uncertainty, respectively. 
With this method, the velocity gradient of the C$^{18}$O emission perpendicular to the outflow is estimated to be 0.27$\pm$0.03 km s$^{-1}$ AU$^{-1}$.

At the outer radii of $\sim$2$\arcsec$ in the P--V diagram along the major axis, the C$^{18}$O emission in the northeast is only observed at blueshifted velocities and only at redshifted velocities in the southwest.
This reveals a velocity gradient with a direction opposite to that in the inner region. 
The presence of an opposite velocity gradient in an outer region to that in an inner region has also been seen in L1527, but on a much larger scale of $\sim$1$\arcmin$ ($\sim$8400 AU; Ohashi et al.~1997; Tobin et al.~2011).
This opposite velocity gradient in the outer region could suggest the presence of counter-rotational motions, as seen in some MHD simulations (e.g., Machida et al.~2014).
Another possibility is that the protostellar envelope in B335 has non-axisymmetric structures on a scale of hundreds of AU, as those observed on larger scales of thousands of AU (e.g., Tobin et al.~2012a).
However, by comparing with the single-dish spectra (Yen et al.~2011), the missing flux is estimated to be more than 90\% around the systemic velocity and more than 50\% at the velocities showing this opposite velocity gradient.
Due to this significant amount of missing flux, the complete velocity structures of the outer emission remain uncertain. 
Future observations providing shorter-spacing data are required to discuss its kinematics and origins.

\subsubsection{Kinematic Model Fitting of Observed P--V Diagrams}
To further estimate the infalling and rotational velocities of the flattened envelope around B335 traced by the C$^{18}$O emission, 
we construct kinematic models of infalling and rotating envelopes and compare them with the observational data. 
The number density ($n$) and temperature ($T$) profiles in the envelope models are assumed to be power-law functions as
\begin{equation}
n(r) = n_0 \cdot (\frac{r}{100\ {\rm AU}})^p, 
\end{equation}
and
\begin{equation}
T(r) = T_0 \cdot (\frac{r}{100\ {\rm AU}})^q.
\end{equation}
The outer radius of the envelope models is adopted to be 3$\arcsec$ (450 AU) to make their diameter twice larger than the maximum angular scale our ALMA observations can sample.
The inclination angle, i.e., the angle between the plane of the sky and the envelope equator, is adopted to be 80$\degr$ (Hirano et al.~1988). 
The number density within 60$\arcdeg$ from the polar axis is artificially set to zero in order to mimic the outflow cavity and the flattened envelope$\footnotemark$. 
The angle of 60$\arcdeg$ is derived from the aspect ratios of the 1.3 mm continuum emission on scales of 40 AU to 500 AU in B335 observed with the SMA and ALMA (Yen et al.~2010; this work), assuming that the lengths of the major and minor axes trace the size scale and thickness of the flattened envelope, respectively. 
The infalling ($V_{\rm in}$) and rotational ($V_{\rm rot}$) motions are assumed to be free fall with a constant angular momentum along the radial direction as
\begin{equation}
V_{\rm in}(r) = \sqrt{\frac{2G(M_*+M_{\rm env})}{r}}, 
\end{equation} 
and 
\begin{equation}
V_{\rm rot}(r) = \frac{j}{r}\sin\theta, 
\end{equation}
where G is the gravitational constant, $M_*$ is the protostellar mass, $M_{\rm env}$ is the envelope mass enclosed within the radius $r$, $j$ is the specific angular momentum of the envelope, and $\theta$ is the angle from the polar axis. 
Since the rotational motion is slow in B335 ($j < 10^{-4}$ km s$^{-1}$ pc; Yen et al.~2010, 2015), 
the expected outer radius of the Keplerian disk (if present) is less than 10 AU; i.e., at least three times smaller than the spatial resolution. 
Therefore, our kinematic model does not include a disk component. 

$\footnotetext{We have also explored the model fitting with the geometrically-thin approximation and a zero density within 30$\arcdeg$ from the polar axis.  We find that the model with the zero density within 60$\arcdeg$ from the polar axis fits the observational results better. Despite the different model assumptions, the best-fit $M_*$ is consistent, 0.05 $M_\sun$, while the best-fit $j$ is comparable and ranges from $\sim$3 $\times$ 10$^{-5}$ to $\sim$5 $\times$ 10$^{-5}$ km s$^{-1}$ pc.}$

The model images are generated by computing the Planck function $B(T)$ and the optical depth for each cell from its $n(r)$ and $T(r)$ on the assumption of the LTE condition and a constant C$^{18}$O abundance of 3.7 $\times$ 10$^{-7}$ (Frerking et al.~1982).
The radiative transfer equation is then integrated along the line of sight (e.g., Yen et a.~2014) in order
to generate a plane-of-sky integrated image.
The model images are processed with the CASA simulator to mimic our observational $uv$ coverage. 
The identical array configuration and observing hour angle are adopted. 
Then, P--V diagrams are generated along the minor and major axes from the simulated model images, 
and we perform $\chi^2$ fitting of the model P--V diagrams to the observed ones.

Our models have four free parameters: $M_*$, $j$, $n_0$, and $T_0$. 
The power-law indices of the density and temperature profiles, $p$ and $q$, are fixed. 
$q$ is adopted to be $-0.4$ (Shirley et al.~2000, 2002). 
Two different values for $p$, $-1.5$ and $-2$, are adopted (Shirley et al.~2002; Harvey et al.~2003a,b; Doty et al.~2010).
We note that with the same $M_*$, $j$, $n_0$, and $T_0$, the models with $p = -2$ always have lower $\chi^2$ compared to those with $p = -1.5$. 
However, due to the simplicity of the geometry of our models, a lower $\chi^2$ with $p = -2.0$ may not simply imply that the density profile of the envelope is really closer to $p = -2.0$, but could mean that the inner envelope is flatter than assumed in our model.
The best-fit parameters are first searched in wider ranges but with coarser steps.  
We then perform the fitting with the parameter ranges of $M_*$ from 0.03 $M_\sun$ to 0.07 $M_\sun$ in steps of 0.01 $M_\sun$, $j$ from 0 to 5 $\times$ 10$^{-5}$ km s$^{-1}$ pc in steps of 1 $\times$ 10$^{-5}$ km s$^{-1}$ pc, $n_0$ from 5 $\times$ 10$^{6}$ cm$^{-3}$ to 4.0 $\times$ 10$^{7}$ cm$^{-3}$ in steps increasing by a factor of two, and $T_0$ from 20 to 60 K in steps of 10 K.   
The final best-fit parameters are $M_*=0.05$ $M_\sun$, $j$ = 2 $\times$ 10$^{-5}$ km s$^{-1}$ pc, $n_0$ = 1$\times$ 10$^{7}$ cm$^{-3}$, and $T_0$ = 30 K.
Figure \ref{c18opv}a and b present the P--V diagrams of our best-fit kinematic model.
We note that this estimated $M_*$ is based on the assumption of the infalling motion being free fall. 
If the infalling velocity is slower than the free-fall velocity, as seen in some other protostars, such as L1527 IRS (Ohashi et al.~2014), L1551 NE (Takakuwa et al.~2013), and L1551 IRS 5 (Chou et al.~2014), 
the estimated $M_*$ is only a lower limit.

The best-fit $M_*$ and $j$ are consistent with those derived using the low-resolution SMA C$^{18}$O (2--1) data (Yen et al.~2010, 2015). 
In a previous study using H$^{13}$CO$^+$ in B335 at an angular resolution of $\sim$5$\arcsec$, 
$M_*$ was estimated to be 0.1 $M_\sun$ (Kurono et al.~2013).
For comparison, we also construct a model with this larger fixed $M_*= 0.1$ $M_\sun$, 
with the remaining parameters the same as our best-fit parameters. 
The P--V diagrams of the model with $M_*=0.1$ $M_\sun$ are shown in Figure \ref{c18opv}c and d.
As a result, our ALMA observations cannot be well explained by the model with $M_*=0.1$ $M_\sun$, 
which shows more high-velocity emission than found in the  observations. 
A kinematic model with a lower $M_*$ of 0.02 $M_\sun$ and best-fit $j$, $\Sigma$, and $T$ is also generated for comparison.
Such a low $M_*$ similarly fails to explain the observed velocity features. 

We note that our fitting is likely biased towards the low-velocity emission whose intensity is a factor of two to three higher than that of the high-velocity emission, 
while the sign of the rotational motion, i.e., the velocity gradient, is clearer at the high velocities.
Therefore, to further investigate the specific angular momentum of the rotational motion responsible for the velocity gradient seen at the high velocities in the P--V diagram along the major axis, 
we have repeated the fitting by excluding the channels at $\Delta V < 1.2$ km s$^{-1}$, fixing $M_*$ at 0.05 $M_\sun$, and increasing $j$ from 0 in steps of 1 $\times$ 10$^{-6}$ km s$^{-1}$ pc. 
We find a best-fit $j$ of 4.3 $\times$ 10$^{-5}$ km s$^{-1}$ pc.
With the same method, a velocity gradient with a magnitude of 0.28 km s$^{-1}$ AU$^{-1}$ consistent to the observed one
is extracted from the simulated model data.
Considering the uncertainty of the velocity gradient measured in the P--V diagram of the C$^{18}$O emission along the major axis, which is $\pm$11\%, 
the $j$ of the protostellar envelope over 180 AU is estimated to be (4.3$\pm$0.5) $\times$ 10$^{-5}$ km s$^{-1}$.
This estimated $j$ is approximately a factor of two below the upper limit of the specific angular momenta on scales of 1000 AU and 100 AU in B335, estimated with the SMA C$^{18}$O (2--1) and CS (7--6) observations, respectively (Yen et al.~2010; 2011).
Because of the low resolution in the previous SMA observations it was impossible to clearly detect this small specific angular momentum.
Therefore, under the assumptions of a free-fall infalling motion and constant angular momentum, 
our results suggest that the flattened envelope on a scale of $\sim$180 AU in B335 traced by the C$^{18}$O (2--1) emission is rotating and infalling toward a 0.05 $M_\sun$ protostar, and its specific angular momentum is estimated to be 4.3 $\times$10$^{-5}$ km s$^{-1}$.

\subsection{Possible Infalling and Rotational Motions Observed in the SO (5$_6$--4$_5$) Emission} 
\begin{figure}
\epsscale{1}
\plotone{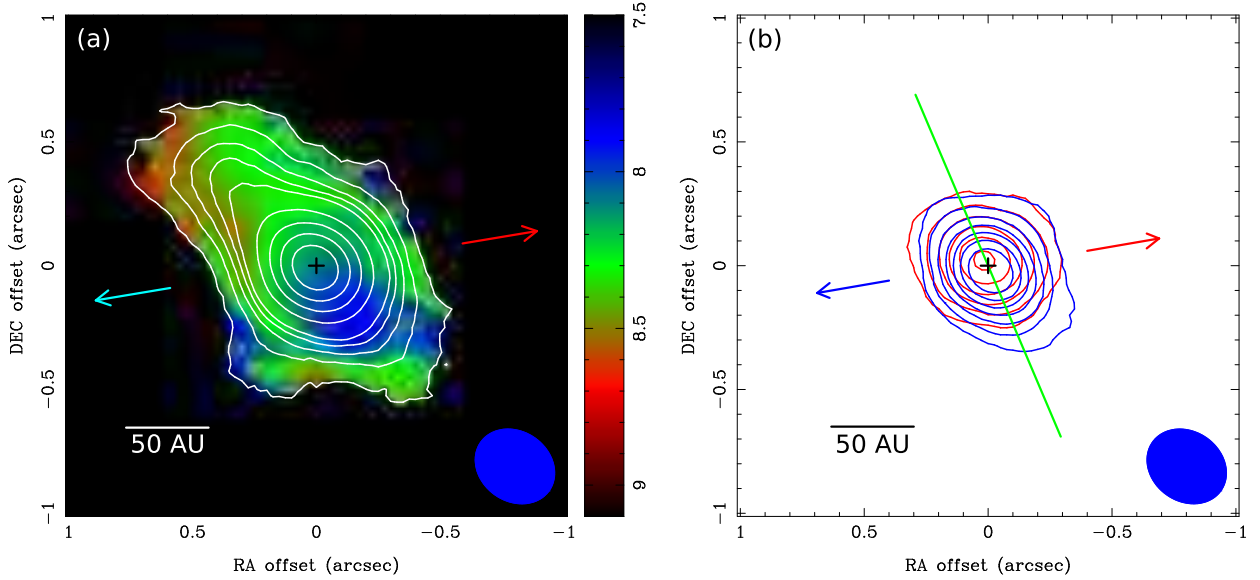}
\caption{(a) Moment 0 map (contours) overlaid on the moment 1 maps (color scale, units in km s$^{-1}$) of the SO (5$_6$--4$_5$) emission in B335. Contour levels are 3$\sigma$, 6$\sigma$, 9$\sigma$, 12$\sigma$, 15$\sigma$, and then in steps of 10$\sigma$, where 1$\sigma$ noise level is 5.1 mJy beam$^{-1}$. (b) Moment 0 map of the high-velocity redshifted and blueshifted SO emission in B335 at $\Delta V > 1.2$ km s$^{-1}$. Contour levels are from 3$\sigma$ in steps of 3$\sigma$, where 1$\sigma$ noise level is 2.7 and 3.1 mJy beam$^{-1}$ in the redshifted and blueshifted emission, respectively. In both panels, crosses indicate the protostellar positions. Filled ellipses at the bottom-right corners present the beam size. Red and blue arrows denote the directions of the red- and blueshifted outflows, respectively. The green line shows the direction of the major axis of the 1.3 mm continuum emission.}
\label{somap}
\end{figure}

SO is one of the major S-bearing molecules in protostellar envelopes. 
Its abundance can be enhanced due to molecular desorption from dust grains when the dust temperature is above $\sim$60 K (Aikawa et al.~2012).
Hence, the presence of the SO emission in protostellar sources is most likely related to warm regions; e.g., due to protostellar heating, outflow shocks, and accretion shocks  (e.g., Spaans et al.~1995; Bachiller \& P\'erez Guti\'errez 1997; Wakelam et al.~2005; van Kempen et al.~2009; Visser et al.~2009; Aota et al.~2015). 
In B335, there are two components seen in the SO emission (Fig.~\ref{somap}). 
One component is more extended ($\sim$1$\arcsec$ or $\sim$150 AU) and elongated along the northeast--southwest direction, 
and the other is compact ($\sim$0\farcs29 or $\sim$44 AU), centered at the protostellar position and has a size and orientation similar to the 1.3 mm continuum emission. 

The extended SO component is unlikely related to protostellar heating or accretion shocks.
The infalling velocity at a radius of 100 AU is $\sim$0.9 km s$^{-1}$. 
Even if there is a dense structure with a density of 10$^9$ cm$^{-3}$ at a radius of $\sim$100 AU for the infalling material to collide, 
the infalling velocity is not high enough to form a shock that can increase the dust temperature to more than 60 K. 
Such an increase would require an infalling velocity of $\gtrsim$2 km s$^{-1}$ (Aota et al.~2015). 
Besides, the temperature on a 100 AU scale is expected to be less than 60 K (e.g., Shirley et al.~2000, 2002).
In contrast to that, the velocity of the jet in B335 seen in the $^{12}$CO line is larger than several tens of km s$^{-1}$. 
At such a high velocity, the temperature can be increased to more than 100 K by shocks even with a pre-shock density of 10$^{4}$ cm$^{-3}$ (e.g., Kaufman et al.~1996).
Besides, the photons from inner accretion disks around protostars can also heat the walls of outflow cavities on a 1000 AU scale to more than 60 K (e.g., Spaans et al.~1995).
Therefore, the extended SO component is more likely related to the outflow or photons escaping from the outflow cavity.

The central compact components of the 1.3 mm continuum and SO emission have similar sizes and orientations, 
suggesting that the central SO component likely also traces the inner flattened structure around the protostar as the 1.3 mm continuum emission does.
In the envelope at a radius of 20 AU, the infalling velocity is estimated to be $\sim$2 km s$^{-1}$, 
which is high enough to form shocks that can rise the dust temperature to more than 60 K (Aota et al.~2015). 
In addition, radiative transfer models of infalling envelopes show that the dust temperature can reach 100 K within a radius of $\sim$25 AU (Visser et al.~2009).
Hence, this component is more likely related to protostellar heating or accretion shocks.
Here, we discuss the velocity structures of this central compact SO component. 

\begin{figure}
\epsscale{1}
\plotone{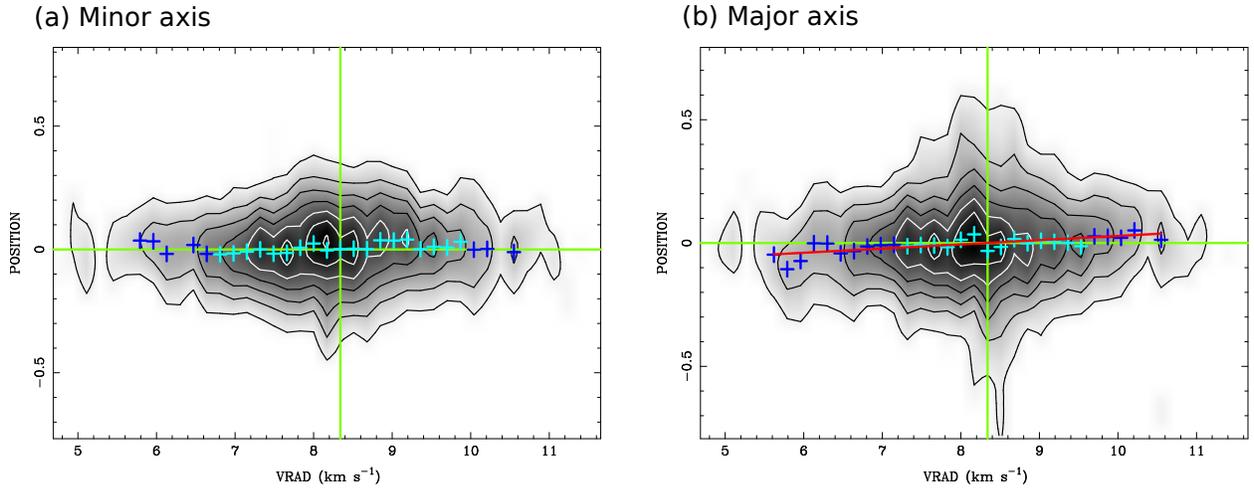}
\caption{P--V diagrams of the SO emission along the minor (left panel) and major (right panel) axis of the 1.3 mm continuum emission. The vertical axis presents position in arcseconds. The horizontal axis is velocity in units of km s$^{-1}$. Blue and light blue crosses present the peak positions measured in different velocity channels in the P--V diagrams. The peak positions shown as blue crosses are used to derive velocity gradients. The red solid line presents the velocity gradient derived from the peak positions. Green vertical and horizontal lines denote the systemic velocity and the protostellar position, respectively. Contour levels are from 3$\sigma$ in steps of 3$\sigma$, where 1$\sigma$ is 4.8 mJy beam$^{-1}$.}
\label{sopv}
\end{figure}

Figure \ref{sopv} presents the P--V diagrams of the SO emission along the major and minor axes of the 1.3 mm continuum emission.  
We derive peak positions in all velocity channels that have emission above 5$\sigma$ in these P--V diagrams (shown as data points). 
From these data points --- identical to the analysis in the previous section --- we find a possible velocity gradient at velocities $\Delta V \lesssim 1.7$ km s$^{-1}$ in the P--V diagram along the minor axis (Fig.~\ref{sopv}a), 
where the western region (offset $>$ 0) is more redshifted and the eastern region (offset $<$ 0) more blueshifted (light blue data points). 
The direction of this velocity gradient is different from that of the infalling motion seen in the C$^{18}$O emission. 
It is more likely associated with the extended SO emission and not originating from the inner flattened structure (Fig.~\ref{somap}). 
At higher velocities, the structures are less clear. 
Including these higher-velocity data points in our analysis, 
we find that there is no detectable velocity gradient along the minor axis in the central compact SO component. 
The absence of a velocity gradient along the minor axis could suggest that there is no clear infalling motion in the inner 10 AU region, 
if the central compact SO component indeed traces the inner flattened structure.

The P--V diagram along the major axis (Fig.~\ref{sopv}b) shows an overall velocity gradient, 
where the northern region (offset $>$ 0) is more redshifted and the southern region (offset $<$ 0) more blueshifted. 
A linear relation $V_{\rm LSR} \propto r$ is fitted to the data points at $\Delta V > 1.2$ km s$^{-1}$ to measure the velocity gradient in the central compact component. 
The data points at the lower velocities (light blue crosses) are excluded to avoid a possible contamination from the extended component. 
The magnitude of the velocity gradient along the major axis is measured to be 0.51$\pm$0.11 km s$^{-1}$ AU$^{-1}$.
Different from that along the minor axis, the presence of the velocity gradient is clear and is not sensitive to the velocity range included in the fitting of the linear relation. 
The direction of this velocity gradient is consistent with the rotational motion on larger scales. 
Hence, a rotational motion is likely present in the inner envelope on a 10 AU scale.

As shown in the P--V diagrams, the velocity structures in the SO emission are not well resolved and only show linear-like features ($V_{\rm LSR} \propto r$). 
In addition, 
the SO emission in protostellar sources is likely related to warm regions, where the excitation conditions can be complicated.
We, therefore, do not further construct kinematic models.
We proceed to estimate the possible rotational velocities from the velocity gradients in the P--V diagrams of the SO emission, assuming that the central compact SO emission traces the inner 10 AU flattened envelope around the protostar.

For estimating the rotational velocity, 
the mean distance of the emission peaks to the protostar in the velocity channels at $\Delta V > 1.2$ km s$^{-1}$ in the P--V diagram along the major axis, 0\farcs031$\pm$0\farcs002 (4.7$\pm$0.3 AU), is adopted as a characteristic radius. 
The rotational velocity is then estimated from the characteristic radius $\times$ the velocity gradient. 
Correction for inclination, 
along the major axis, the velocity needs to be divided by $\sin80\degr$, while the radius does not. 
The rotational velocity at a radius of 4.7$\pm$0.3 AU is estimated to be 2.7$\pm$0.5 km s$^{-1}$. 
This corresponds to a specific angular momentum of (5.4$\pm$1.1) $\times$ 10$^{-5}$ km s$^{-1}$ pc. 
For comparison, we also estimate the rotational velocity using the moment 0 map of the high-velocity SO emission (Fig.~\ref{somap}b). 
By adopting one half of the separation between the high-velocity blueshifted and redshifted components as the characteristic radius, 3.4$\pm$0.3 AU, 
the rotational velocity at that radius is estimated to be $\sim$2.0 km s$^{-1}$ from the velocity gradient measured in the moment 0 map.
This corresponds to a specific angular momentum of 3.3 $\times$ 10$^{-5}$ km s$^{-1}$ pc. 
Thus, the specific angular momenta estimated from the P--V diagram along the major axis and from the moment 0 map of the high-velocity components are comparable. 
The difference is likely due to (1) the different position angles of the velocity gradient in the moment 0 map ($\sim$30$\arcdeg$) and the major axis of the continuum emission ($\sim$14$\arcdeg$) and (2) the different methods to compute characteristic radii. 
In the moment 0 map, the characteristic radius is derived from the peak position of the integrated emission, 
while that in the P--V diagram is from the mean of the peak positions in each velocity channel. 
We can, thus, conclude that 
the specific angular momentum on a 10 AU scale estimated from the SO results, $\sim$(3--5) $\times$ 10$^{-5}$ km s$^{-1}$ pc, is comparable to that on a 100 AU scale from the C$^{18}$O results ($\sim$5 $\times$ 10$^{-5}$ km s$^{-1}$ pc).

Since there is no clear sign of the infalling motion,
there is a possibility that the rotational motion can be more dominant than the infalling motion within a 10 AU scale around the protostar. 
The presence of a rotational motion without a clear infalling motion is an observational signature of a Keplerian disk. 
Assuming the central protostellar mass is 0.05 $M_\sun$, as estimated from the C$^{18}$O results, 
the expected Keplerian velocity at a radius of 4.7 AU is $\sim$3.1 km s$^{-1}$, and that at 3.4 AU is $\sim$3.6 km s$^{-1}$. 
From the SO P--V digram along the major axis, the rotational velocity at a radius of 4.7 AU is estimated to be $\sim$2.7$\pm$0.5 km s$^{-1}$, 
and that estimated from the moment 0 of the high-velocity SO emission is $\sim$2.0 km s$^{-1}$ at a radius of 3.4 AU. 
Hence, the rotational velocity on the innermost 10 AU scale is likely smaller than the Keplerian velocity around a 0.05 $M_\sun$ protostar.  
Assuming the specific angular momentum within the 10 AU scale is constant with $\sim$(3.3--5.4) $\times$ 10$^{-5}$ km s$^{-1}$ pc and the protostellar mass is 0.05 $M_\sun$, 
the expected radius of the Keplerian disk around B335, $j^2/GM_*$, is derived to be 1--3 AU, where $j$ is the specific angular momentum.
Note that the estimated protostellar mass from the C$^{18}$O results can be a lower limit, 
and the expected Keplerian velocities at these radii can be larger, leading to an even smaller disk size.

\section{Discussion}
Our ALMA results and kinematic models suggest that in B335 the protostellar envelope within 180 AU traced by the C$^{18}$O emission is likely infalling and rotating. 
Under the assumptions of a free-fall infalling motion and a constant angular momentum, 
the protostellar mass is estimated to be 0.05 $M_\sun$.
The specific angular momentum of the infalling and rotating envelope is $\sim$4.3 $\times$ 10$^{-5}$ km s$^{-1}$ pc. 
The innermost region of 40 AU traced by the SO emission is likely rotating with a similar specific angular momentum, 
without showing a clear sign of infalling motion.
These results suggest that the gas motion could change from an infall-dominated to a rotation-dominated motion in the innermost region. 
The rotational velocity of the SO component is likely smaller than the expected Keplerian velocity around a protostar of 0.05 $M_\sun$.
This suggests the absence of a compact Keplerian disk with a size of tens of AU.
Assuming the specific angular momentum within the 10 AU scale is $\sim$(3--5) $\times$ 10$^{-5}$ km s$^{-1}$ pc, 
the expected radius of the Keplerian disk around B335 is derived to be 1--3 AU. 
This disk radius is one to two orders of magnitude smaller than what is observed in other Class 0 protostars, e.g., $\sim$150 AU in VLA 1623A (Murillo et al.~2013), $\sim$50 AU in L1527 IRS (Ohashi et al.~2014), and $\sim$120 AU in HH 212 (Lee et al.~2014).

There are two possible scenarios to explain such a small Keplerian disk size in B335: (1) magnetic braking and (2) young age. 
On one hand,  
if magnetic braking is efficient in B335, this mechanism can remove angular momentum from the infalling material and suppress the disk size. 
On the other hand, 
if B335 is much younger than those Class 0 protostars exhibiting Keplerian disks with sizes of tens of AU, there is less angular momentum which has been transferred to the innermost region, 
and the Keplerian disk has not yet accumulated sufficient angular momentum to grow in size. 
Below, we first describe the infalling and rotational motions from 0.1 pc to the inner 100 AU scales in B335,
and we then discuss the two scenarios in more detail.

\subsection{Gas Motion from 0.1 pc to Inner 100 AU Scales}
\subsubsection{Infalling Motion}
In B335, the infalling motion on a larger scale of a few thousand AU is detected in single-dish observations in the CS and H$_2$CO lines at angular resolutions of $\sim$11\arcsec--28$\arcsec$ (Zhou et al.~1993, 1995; Choi et al.~1995). 
By comparing the observed spectra with theoretical models of an inside-out collapse, 
the radius of the infalling region in B335 is estimated to be $\sim$3700--4400 AU
with a central protostellar mass of $\sim$0.22--0.26 $M_\sun$. 
Unlike the single-dish observations,  
the interferometric results of B335 show that the CS (5--4) emission on a $\sim$10$\arcsec$ ($\sim$1500 AU) scale primarily traces the outflow rather than the infalling gas (Wilner et al.~2000).
Indeed, signs of outflow contamination are possibly seen in those single-dish CS spectra, where the emission at higher velocities ($\Delta V \gtrsim 1$ km s$^{-1}$) is more intense than model predictions of an inside-out collapse. 
Hence, the infalling radius from modelling of the single-dish spectra can be overestimated due to the outflow contamination. 
It is, however, not straightforward to quantify this effect.

The inside-out collapse model predicts a density profile that changes from having a power-law index of $-2$ to $-1.5$ at the infalling radius (Shu 1977). 
The presence of such a change in the density profile in B335 was suggested by Harvey et al.~(2001) from the near-infrared extinction observed with the Keck telescope and by Kurono et al.~(2013) from the intensity distribution of the H$^{13}$CO$^+$ (1--0) emission observed with the Nobeyama 45m telescope and the Nobeyama Millimeter Array (NMA). 
These observations find that in B335 the density profile in the inner region is approximately $\propto r^{-1.5}$ and that in the outer region is $\propto r^{-2}$, 
with a turning point radius at $\sim$3900--4000 AU. 
Therefore, 
all these analyses of density profiles and single-dish spectra suggest a consistent infalling radius of $\sim$4000 AU, leading to a protostellar mass of $\sim$0.24 $M_\sun$.  
On the contrary to this, 
the SCUBA 850 $\mu$m and 450 $\mu$m data do not support the presence of a turning point in the density profile at a radius of $\sim$4000 AU, and the power-law index of the density profile within a radius of $\sim$50$\arcsec$ ($\sim$7500 AU) is estimated to be $-1.8$ (Shirley et al.~2002). 
The SCUBA results further suggest that, if the infalling radius is indeed present, it is likely at $\sim$4$\arcsec$ ($\sim$600 AU; Shirley et al.~2002), although this is more than two times smaller than the angular resolution of the SCUBA observations ($\sim$8\arcsec--15\arcsec or 1200--2250 AU). 
In contrast to the SCUBA results,  
with the PdBI 1.3 mm and 2.7 mm continuum observations at subarcsecond resolutions, the power-law index of the density profile within a radius of 20$\arcsec$ (3000 AU) is estimated to be approximately $-1.5$ (Harvey et al.~2003a,b).  
A later study combining the SCUBA data, the spectral energy distribution, and the infrared extinction suggests that the power-law index of the density profile in B335 is $\sim$1.5--1.9 at radii between 10$\arcsec$ (1500 AU) and 50$\arcsec$ (7500 AU), 
and there is no clear sign of a turning point in the density profile (Doty et al.~2010). 
In summary, a comparison of all these results$\footnotemark$ suggests that the size of the collapsing region in B335 is likely on a scale of $\sim$1000 or up to a few thousand AU.

\footnotetext{Zhou et al.~(1993), Choi et al.~(1995), Shirely et al.~(2002), and Harvey et al.~(2003a,b) adopted a distance of 250 pc to B335 in their analyses. The infalling radius and the derived protostellar mass shown here are scaled from the distance of 250 pc to 150 pc. Doty et al.~(2010) adopted two distances, 250 pc and 150 pc, in their analyses.}

Besides the observational signatures of the infalling motion in the single-dish spectra and the density profile, 
the velocity structures of the infalling motion on a scale of a few thousand AU are spatially resolved with a combination of  observations in the H$^{13}$CO$^+$ (1--0) line with the Nobeyama 45-m telescope and NMA, resulting in an angular resolution of $\sim$5$\arcsec$ (Kurono et al.~2013).
They find that the velocity structures can be best explained with a free-falling motion toward a protostar of 0.1 $M_\sun$. 
If a protostellar mass of 0.05 $M_\sun$ or 0.15 $M_\sun$ is adopted in their model, 
narrower or wider line widths are expected at a radius of $\lesssim$1000 AU ($\sim$7\arcsec), respectively (Kurono et al.~2013). 
From the gas motions on a next smaller scale within a few hundred AU --- probed by our SMA and ALMA observations with the highest angular resolutions among these observations --- 
the protostellar mass is estimated to be 0.05 $M_\sun$ and unlikely as large as 0.1 $M_\sun$ (Fig.~\ref{c18opv}). 
The difference between the ALMA and Nobeyama results could be reconciled if the self-gravity of the envelope material within a radius of 1000 AU, having a mass of $\sim$0.05 $M_\sun$ (Kurono et al.~2013), was taken into account in their analyses. 
However, the protostellar mass estimates from the single-dish spectra by Zhou et al.~(1993, 1995) and Choi et al.~(1995) are almost a factor of five larger than ours, 
although self-gravity of the collapsing material is considered in their models. 
These results show a tendency that the protostellar mass is lower when estimated from observations probing smaller scales.

\subsubsection{Rotational Motion}
B335 is associated with extended structures on a 0.4 pc scale. 
On an inner 0.2 pc scale, the structures appear to be flattened and become more symmetric with respect to the outflow axis (Launhardt et al.~2013). 
The associated dense core within the 0.2 pc scale exhibits a smooth velocity gradient, perpendicular to the outflow axis,  
which can be described with a linear relation, i.e., $V_{\rm LSR} \propto R$ (Saito et al.~1999).  
This velocity gradient can be a signature of large-scale rotational motion (e.g., Goodman et al.~1993) or infalling motion (e.g., Tobin et al.~2012a; Peretto et al.~2014).  
Large-scale infalling motions driven by the gravity of a central protostar tend to show higher velocities at smaller radii (e.g., Tobin et al.~2012a). This is different from the velocity features observed in B335. 
Although large-scale inflows along filaments can also show a similar linear velocity gradient (e.g., Peretto et al.~2014), 
these inflows are expected to appear in filaments with a more uniform density (e.g., Pon et al.~2012). 
The central 0.2 pc region in B335 has already shown a centrally-peaked density distribution (e.g., Saito et al.~1999; Harvey et al.~2001). 
Therefore, the velocity gradient on the 0.2 pc scale in B335 is more likely due to a rotational motion of the associated dense core. 

These large-scale velocity gradients perpendicular to the outflow axis in B335 were measured to be $\sim$0.6--1.0 km s$^{-1}$ pc$^{-1}$ over a 20,000 AU scale (Saito et al.~1999; Kurono et al.~2013), and $\sim$0.8 km s$^{-1}$ pc$^{-1}$ over a 9000 AU scale (Yen et al.~2011). 
Assuming that the dense core is rotating as a rigid body, 
its specific angular momentum can be derived from the observed velocity gradients perpendicular to the outflow.
It is estimated to be (5.6--9.4) $\times$ 10$^{-3}$ km s$^{-1}$ pc at a radius of 20,000 AU and 1.5 $\times$ 10$^{-3}$ km s$^{-1}$ pc at a radius of 9000 AU.  
Note that the large-scale specific angular momentum can be overestimated, as demonstrated by Dib et al.~(2010).
On a 1000 AU scale, 
the SMA results suggest that the upper limit of the specific angular momentum is 7 $\times$ 10$^{-5}$ km s$^{-1}$ pc. 
With our ALMA observations, the specific angular momentum within a 100 AU scale is estimated to be $\sim$(3--5) $\times$ 10$^{-5}$ km s$^{-1}$ pc.

\begin{figure}
\epsscale{1}
\plotone{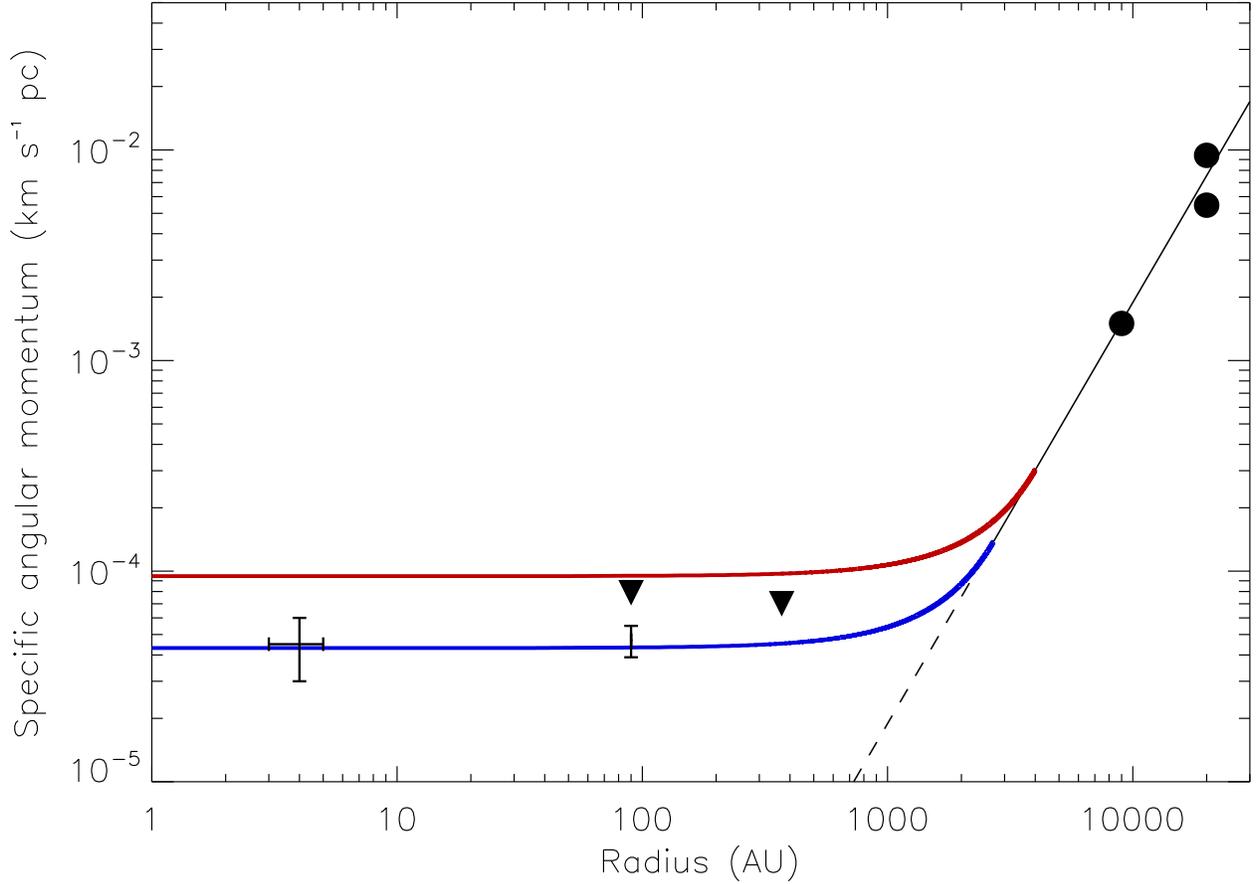}
\caption{Specific angular momentum as a function of radius in B335. Circles are measurements from singe-dish observations (Saito et al.~1999; Yen et al.~2011; Kuronu et al.~2013). The triangles denote the upper limits estimated from SMA observations (Yen et al.~2010; 2011). Data points with error bars are the measurements from the ALMA observations in the C$^{18}$O and SO emission (this work). Red and blue curves present the expected profiles of a rotating dense core undergoing an inside-out collapse at the time that its infalling radius is 4000 AU and 2700 AU, respectively. The initial rotational motion of the dense core is assumed to be a rigid-body rotation with an angular velocity of 2.6 $\times$ 10$^{-14}$ s$^{-1}$, shown as a dashed line.}
\label{jr}
\end{figure}

Figure \ref{jr} presents the specific angular momentum as a function of radius in B335. 
The values fall quickly as the radius decreases from 20,000 AU to the inner 1000 AU. 
Such a steep profile 
is seen in statistical studies of the relation between dense core sizes and their angular momenta
with a power-law index of $1.6$ (Goodman et al.~1993). 
Within a radius of 1000 AU, the profile of the specific angular momentum in B335 becomes flat. 
A change from a steep to a flat profile,  as the radius decreases,  is expected in the inside-out collapse model with rotational motion. 
Assuming that initially the dense core is rotating as a rigid body and, hence, exhibits a steep specific-angular-momentum profile, 
material with a larger specific angular momentum at an outer radius starts to fall in and accelerate to a supersonic speed
as the expansion wave propagates outward.
If the angular momentum of the infalling material is conserved, 
the infalling material carries its angular momentum inward, 
and its supersonic motion makes the profile of the specific angular momentum within the infalling radius become flat and almost constant. 
Outside the infalling radius, 
the material remains static, 
and the profile of the specific angular momentum does not change. 
Figure \ref{jr} shows that the behavior of the radial profile of the specific angular momentum in B335 is generally consistent with the scenario of an inside-out collapse of a rotating dense core.
In this scenario, 
the expected angular momentum on a 100 AU scale can be estimated from the infalling radius and the large-scale rotational motion in B335, under the assumption that the angular momentum of the infalling material is conserved. 

\subsection{Hints of Magnetic Braking?}

The protostellar mass in B335 is estimated from single-dish spectra (Zhou et al.~1993, 1995; Choi et al.~1995), the Nobeyama results (Kuruno et al.~2013), and our SMA and ALMA results (Yen et al.~2010, 2015; this work). 
These estimates are all based on the velocity features of the infalling motion but not on resolved Keplerian rotation. 
Therefore, the estimated protostellar masses more likely present the corresponding gravity that pulls gas toward the center instead of the genuine protostellar mass. 
The comparison shows that the estimated protostellar masses decrease with spatial scales. 
This can hint that the net gravity pulling gas inward decreases in the inner regions. 
In other words, the infalling velocity on a smaller scale can be more significantly slower than the free-fall velocity$\footnotemark$, leading to a smaller estimated protostellar mass.
Possible explanations for this are contributions such as turbulence, thermal pressure, magnetic pressure, or magnetic tension, that become more significant and work against gravity. 
Furthermore, 
in the studies of the single-dish spectra and the Nobeyama results, the infalling radius in B335 is estimated to be $\sim$4000 AU (Zhou et al.1993; Zhou 1995; Choi et al.~1995; Kuruno et al.~2013). 
With the observed large-scale rotation (2.6 $\times$ 10$^{-14}$ rad s$^{-1}$; the mean value of single-dish measurements) and the infalling radius of $\sim$4000 AU, 
the expected angular momentum on a 100 AU scale is estimated to be $\sim$9.5 $\times$ 10$^{-5}$ km s$^{-1}$ pc, assuming the angular momentum of the infalling material is conserved.
Interestingly,  
the specific angular momentum within a 100 AU scale measured from our ALMA observations, $\sim$(3--5) $\times$ 10$^{-5}$ km s$^{-1}$ pc, is approximately a factor of two to three lower than the expectation, if the large-scale specific angular momentum is not overestimated. 
Our results, thus, possibly hint that the angular momentum of the infalling material on a 1000 AU scale in B335 is partially removed. 

\footnotetext{In this scenario, the protostellar mass can be as large as $\sim$0.22--0.26 $M_\sun$ (Zhou et al.~1993; Zhou 1995; Choi et al.~1995). That is comparable to the envelope mass within a radius of 4000 AU (e.g., Harvey et al.~2003a,b), and becomes more dominant on smaller scales. Therefore, the expected free-fall velocity increases with the radius decreases, similar to the solution of the inside-out collapse model (Shu 1977).}

The comparison of the gas motions on scales from 0.1 pc to the inner 100 AU hints that the net force pulling gas inward possibly decreases in the inner region, and the angular momentum of the infalling material is likely partially removed during the infalling motion. 
Both phenomena are shown in MHD simulations of collapsing dense cores (e.g., Mellon \& Li 2008, 2009; Li et al.~2011). 
As the magnetic field is dragged toward the inner region with the infalling motion, 
the magnetic field tension and pressure become more important and can slow down the acceleration of the infalling material. 
In addition, the thermal pressure can also be enhanced in the inner region, if there is substantial material collapsing vertically along the magnetic field to the equator (e.g., Mellon \& Li 2008).  
Observationally, such infalling velocities slower than the free-fall velocities by 30\% to 50\% have been seen in several protostellar sources, such as L1527 IRS (Ohashi et al.~2014), L1551 NE (Takakuwa et al.~2013), and L1551 IRS 5 (Chou et al.~2014). 
Besides the infalling motion, 
the rotational motion can also be affected by the magnetic field. 
As the infalling material rotates around the protostar, 
the magnetic field can be twisted.  
The twisted field can then remove angular momentum of the infalling material and transfer it outward, leading to the so-called ``magnetic braking''. 
Although (isotropic) turbulent or thermal pressure can also work against gravity, 
it is unlikely that 
they develop a preferential direction to slow down the azimuthal motion (i.e., rotational motion).
Therefore, our ALMA results could suggest that the gas motions in B335 are slowed down by the magnetic field.

If the difference between the expected and observed specific angular momentum at a radius of $\sim$100 AU is indeed caused by magnetic braking, we can estimate the order of magnitude of the required magnetic field strength. 
Assuming the system has reached steady state (i.e., $\partial j/\partial t = 0$), 
the change in angular momentum can be approximated as
\begin{equation}
V_{\rm in} \frac{\Delta j}{\Delta r} = \frac{r B_z B_\phi}{2\pi \Sigma},
\end{equation}
where $V_{\rm in}$ is the infalling velocity, $\Delta j$ is the amount of angular momentum removed, $\Delta r$ is the distance 
over which the infalling material has been slowed down due to magnetic braking of the vertical and azimuthal field components $B_z$ and $B_\phi$
(Krasnopolsky \& K\"onigl 2002).
As estimated from our kinematic models for the C$^{18}$O emission, at $r = 100$ AU, 
$V_{\rm in}$ is $\sim$1 km s$^{-1}$, $\Sigma$ is 2 $\times$ 10$^{22}$ cm$^{-2}$. $\Delta j$ at $r = 100$ AU is $\sim$ 5.5 $\times$ 10$^{-5}$ km s$^{-1}$ pc (Figure \ref{jr}).  
For simplicity, we assume $B_z \sim B_\phi$ and $\Delta r \sim 1000$ AU
which corresponds to the radius where the expected specific angular momentum profile flattens to an inner radius of 100 AU.
The resulting magnetic field strength for magnetic braking in B335 is then estimated to be $\sim$200 $\mu$G.
This value is similar to the Zeeman measurements of molecular clouds at column densities of 10$^7$ cm$^{-3}$ (e.g., Crutcher 2012). 

The MHD simulations suggest that the efficiency of magnetic braking can be related to the orientations of the magnetic field and rotational axes of dense cores (e.g., Hennebelle \& Ciardi 2009; Joos et al.~2012; Li et al.~2013). 
If the magnetic field and rotational axis 
are aligned, 
magnetic braking is more efficient in removing the angular momentum of the infalling material.
The formation of large-scale ($>$10 AU) Keplerian disks can, thus, be suppressed.
The orientations of the magnetic field from large to small scales in B335 are observed. 
Near-infrared polarimetric observations$\footnotemark$ show that the magnetic field on a 0.2 pc scale has a mean position angle of $\sim$104$\degr$ and is aligned with the outflow axis within 5$\degr$ (Bertrang et al.~2014). 
On a scale of thousands of AU, the magnetic field shows more disordered structures with a mean position angle of $\sim$18$\degr$, placing it almost perpendicularly to the outflow, as revealed by single-dish polarimetric observations at submillimeter wavelengths (Wolf et al.~2003; Chapman et al.~2013; Hull et al.~2014). 
On a smaller scale of $\sim$1000 AU, the orientation of the magnetic field becomes again more aligned with the outflow with a mean position angle of $\sim$123$\degr$, as observed with interferometric polarimetric observations at millimeter wavelengths (Hull et al.~2014). 
However, the signal-to-noise ratios of these polarization detections on a 1000 AU scale are all less than 3$\sigma$.
With these marginal detections the magnetic field orientations on this scale are possibly still uncertain.
The outflow axis can be considered as the rotational axis. 
Hence, B335 possibly shows all signs of magnetic braking, i.e., no sign of a Keplerian disk larger than 10 AU and a magnetic field aligned with the rotational axis in its dense core. This is all consistent with expectations from MHD simulations. 
Nevertheless, these polarimetric observations also indicate that the mean orientations of the magnetic field change in a complicated manner from 0.2 pc to 1000 AU scales. 
Synthetic polarized continuum images of MHD simulations of collapsing dense cores show that the observed mean magnetic field orientation can depend on which part of the magnetic field structures is sampled by observations (e.g., Davidson et al.~2014).
Therefore, a more detailed comparison between magnetic field orientations from large to small scales in B335 and those from synthetic polarization continuum images computed by MHD simulations of efficient magnetic braking, like the study of L1527 by Davidson et al.~(2014), is essential to further investigate and conclude on the role of the magnetic field in B335.

\footnotetext{The near-infrared polarization is caused by dust grains absorbing light from background stars. Hence, the observed polarization could be contaminated by foreground dust grains which are not associated with B335.} 

\subsection{Hints of Young Age?}
The studies of the density profiles in B335 suggest that the size of the collapsing region is likely on a scale of $\sim$1000 AU or up to a few thousand AU (Harvey et al.~2001, 2003a,b; Shirley et al.~2002; Doty et al.~2010; Kurono et al.~2013). 
If the infalling radius in B335 is only $\sim$2700 AU but not $\sim$4000 AU as in the scenario of magnetic braking,  
the observed specific angular momentum on a 100 AU scale can still be consistent with expectations of an inside-out collapse model without further incorporating the effect of a magnetic field,
assuming that the large-scale specific angular momentum is 1.5 $\times$ 10$^{-3}$ km s$^{-1}$ pc at a radius of 9000 AU as measured with the SMT observations (Yen et al.~2010).
In this scenario, the angular momentum of the infalling material is conserved. 
Since only a smaller amount of angular momentum is transferred to and accumulated in the inner region, a Keplerian disk $>$10 AU is not yet formed. 
The size of the Keplerian disk is expected to increase with time, as more angular momentum is transferred to the inner region with proceeding collapse. 

In this scenario, the time scale of the collapse in B335 is $\sim$5.3 $\times$ 10$^4$ yr, assuming a sound speed of 0.23 km s$^{-1}$ (e.g., Zhou et al.~1993). 
This time scale is comparable to the dynamical time scale of the outflow and the time scale of accretion to form a 0.05 $M_\sun$ protostar in B335. 
In B335, the outflow extends over $\sim$8$\arcmin$ ($\sim$0.35 pc) along the east--west direction. 
The western lobe of the outflow has broken through the associated dense core, and its size of $\sim$0.17 pc is likely only a lower limit (e.g., Hirano et al.~1988). 
A possible sign of the eastern lobe of the outflow is seen at $\sim$10$\arcmin$ ($\sim$0.44 pc) away from the protostar (Langer et al.~1986). 
The mass-weighted mean velocity of the outflow is $\sim$13 km s$^{-1}$ after correcting for its inclination angle (Hirano et al.~1988).
Hence, the dynamical time scale of the outflow is likely longer than 1.3 $\times$ 10$^4$ yr, or even up to 3.3 $\times$ 10$^4$ yr if the outflow indeed extends to 10$\arcmin$. 
The bolometric luminosity of B335 is $\sim$1.5 $L_\sun$. 
Assuming the entire luminosity to result from the gravitational energy released by the material accreted onto the surface of the protostar, 
the mass accretion rate ($\dot{M}_{\rm acc}$) can be estimated to be $\sim$3.0 $\times$ 10$^{-6}$ $M_\sun$ yr$^{-1}$ from 
\begin{equation}
\dot{M}_{\rm acc} = \frac{L_{\rm bol} R_*}{GM_*}, 
\end{equation}
where $R_*$ is the protostellar radius, assumed to be 3 $R_\sun$ (Stahler et al.~1980). 
The accretion time scale can be estimated as $M_*/\dot{M}_{\rm acc}$ to be $\sim$1.7 $\times$ 10$^4$ yr, assuming the accretion rate is constant during the collapse
and neglecting possible episodic accretion activities (e.g., Dunham \& Vorobyov 2012).
Considering uncertainties in the sound speed, the outflow velocity, and the variability of the accretion rate, all these time scales appear consistent with each other. 
Furthermore, 
this scenario can also reduce the discrepancy in the protostellar masses estimated from the single-dish spectra and our SMA and ALMA results. 
If the infalling radius is overestimated in the single-dish studies, 
their protostellar masses are also overestimated.  
Therefore, B335 could be younger than other Class 0 protostars exhibiting Keplerian disks with sizes of tens of AU, and consequently, a large-scale Keplerian disk is not yet formed in B335.

In summary, 
if the size of the collapsing region in B335 is larger, 
a larger amount of angular momentum is expected to be transferred to the inner 100 AU scale. 
In this case, 
the scenario of magnetic braking, which slows down both the infalling and rotational motions, 
would be favoured to reconcile the discrepancy between the specific angular momentum and protostellar mass estimated with our ALMA observations and those with the single-dish spectra by Zhou et al.~(1993), Zhou (1995), and Choi et al.~(1995) and the density profile by Harvey et al.~(2001, 2003a,b) and Kuruno et al.~(2013). 
On the other hand, 
if the size of the collapsing region is smaller (hints by Wilner et al.~2000, Shirley et al.~2002, and Doty et al.~2010), 
the absence of a large-scale Keplerian disk and the low protostellar mass estimated with our ALMA observations are likely the consequence of B335 being younger than other Class 0 protostars that exhibit Keplerian disks with sizes of tens of AU. 
Hence, only a small amount of the angular momentum has been transferred to the inner 100 AU scale, which is insufficient to form a large-scale Keplerian disk. 
Observations that spatially resolve the velocity structures on a scale of a few thousand AU or spectra at high angular resolutions that reveal self-absorption without outflow contamination can help to constrain the size of the collapsing region. 
In addition, future studies comparing both the gas motion and the magnetic field structures from large to small scale with MHD simulations with and without efficient magnetic braking are essential to assess the importance of the magnetic field.

\section{Summary}
We present ALMA cycle 2 observations in the 1.3 mm continuum and the C$^{18}$O (2--1) and SO (5$_6$--4$_5$) lines at an angular resolution of $\sim$0\farcs3 toward the Class 0 protostar B335. We estimate the infalling and rotational velocities of its protostellar envelope from 100 AU to 10 AU scales. By comparing our and previous observational results, we study the gas motions from the scale of the associated dense core ($\sim$0.1 pc) to the inner 10 AU region.  This allows us to quantify net gravity and angular momentum transfer from large to small scales. Our main results are summarized here. 
\begin{enumerate}
\item{Central compact components with sizes of $\sim$40--180 AU are detected in the 1.3 mm continuum, C$^{18}$O, and SO emission. 
These central components are elongated perpendicularly to the outflow axis and most likely trace the inner flattened envelope. 
The central component of the C$^{18}$O emission over 180 AU shows an overall velocity gradient along the east--west direction, i.e., along the outflow axis.  
In addition, a smaller velocity gradient perpendicular to the outflow axis is also seen in the central component at higher velocities ($\Delta > 1.2$ km s$^{-1}$). 
These velocity gradients are likely the signs of infalling and rotational motions in the protostellar envelope, 
with the infalling motion being dominant over the rotational motion. 
In contrast to that, the central SO component over 40 AU shows a velocity gradient perpendicular to the outflow axis without a sign of a velocity gradient along the outflow axis.
The SO results likely suggest the presence of a rotational motion without a dominant infalling motion in the innermost envelope.}

\item{Kinematic models of infalling and rotating envelopes are constructed to fit the velocity features seen in the P--V diagrams of the C$^{18}$O emission along major and minor axes of the 1.3 mm continuum emission. 
Assuming the infalling motion being free fall and a constant specific angular momentum, 
the protostellar mass and the specific angular momentum are estimated to be 0.05 $M_\sun$ and 4.3$\pm$0.5 $\times$ 10$^{-5}$ km s$^{-1}$ pc, respectively.
We also measure the rotational motion from the velocity gradients in the SO emission, 
and the specific angular momentum within a 10 AU scale is estimated to be (3--5) $\times$ 10$^{-5}$ km s$^{-1}$ pc, 
comparable to that of the C$^{18}$O emission over 180 AU.   
Comparing the estimated protostellar mass and specific angular momentum,
we find that the rotational motion has not yet reached Keplerian velocity, 
and no signature of a Keplerian disk is observed even on a 10 AU scale.
The expected radius of the Keplerian disk in B335 is derived to be 1--3 AU.}

\item{There are two scenarios to explain the low specific angular momentum within a 100 AU scale and the small Keplerian disk in B335, as compared to other Class 0 protostars with observed Keplerian disks: (1) magnetic braking and (2) young age. 
If the infalling radius in B335 is $\sim$4000 AU ($\sim$27\arcsec), as suggested by previous studies of infall signatures in single-dish spectra, 
the specific angular momentum within a 100 AU scale observed with ALMA is a factor of two lower than the expectation from the inside-out collapse model with a conserved angular momentum. 
In addition, the protostellar mass estimated from the observed infalling motion within a 100 AU scale is also a factor of four lower than that predicted by the inside-out collapse model. 
These discrepancies could hint that the infalling and rotational motions are slowed down by the magnetic field, 
and the disk growth is suppressed. 
The required field strength to induce this magnetic braking is estimated to be $\sim$200 $\mu$G, 
On the contrary, if the infalling radius is overestimated and is only $\sim$2700 AU ($\sim$18\arcsec), 
the observed specific angular momentum within a 100 AU scale is consistent with the expectation from the inside-out collapse model with a conserved angular momentum. 
In this scenario, the collapse time of B335 is estimated to be 5.3 $\times$ 10$^4$ yr, comparable to its dynamical time scale of the outflow and the accretion time scale, 
and a large Keplerian disk can form with the proceeding collapse.}

\end{enumerate} 

\acknowledgments
This paper makes use of the following ALMA data: ADS/JAO.ALMA\#2013.1.00879.S. ALMA is a partnership of ESO (representing its member states), NSF (USA) and NINS (Japan), together with NRC (Canada) and NSC and ASIAA (Taiwan), in cooperation with the Republic of Chile. The Joint ALMA Observatory is operated by ESO, AUI/NRAO and NAOJ. We thank all the ALMA staff supporting this work. S.T. acknowledges a grant from the Ministry of Science and Technology (MOST) of Taiwan (MOST 102-2119-M-001-012-MY3) in support of this work. P.M.K. acknowledges support from MOST 103-2119-M-001-009 and Academia Sinica Career Development Award. 

\appendix

\section{Velocity Channel Maps of the C$^{18}$O and SO Emission}\label{chan}
\begin{figure}
\epsscale{1}
\plotone{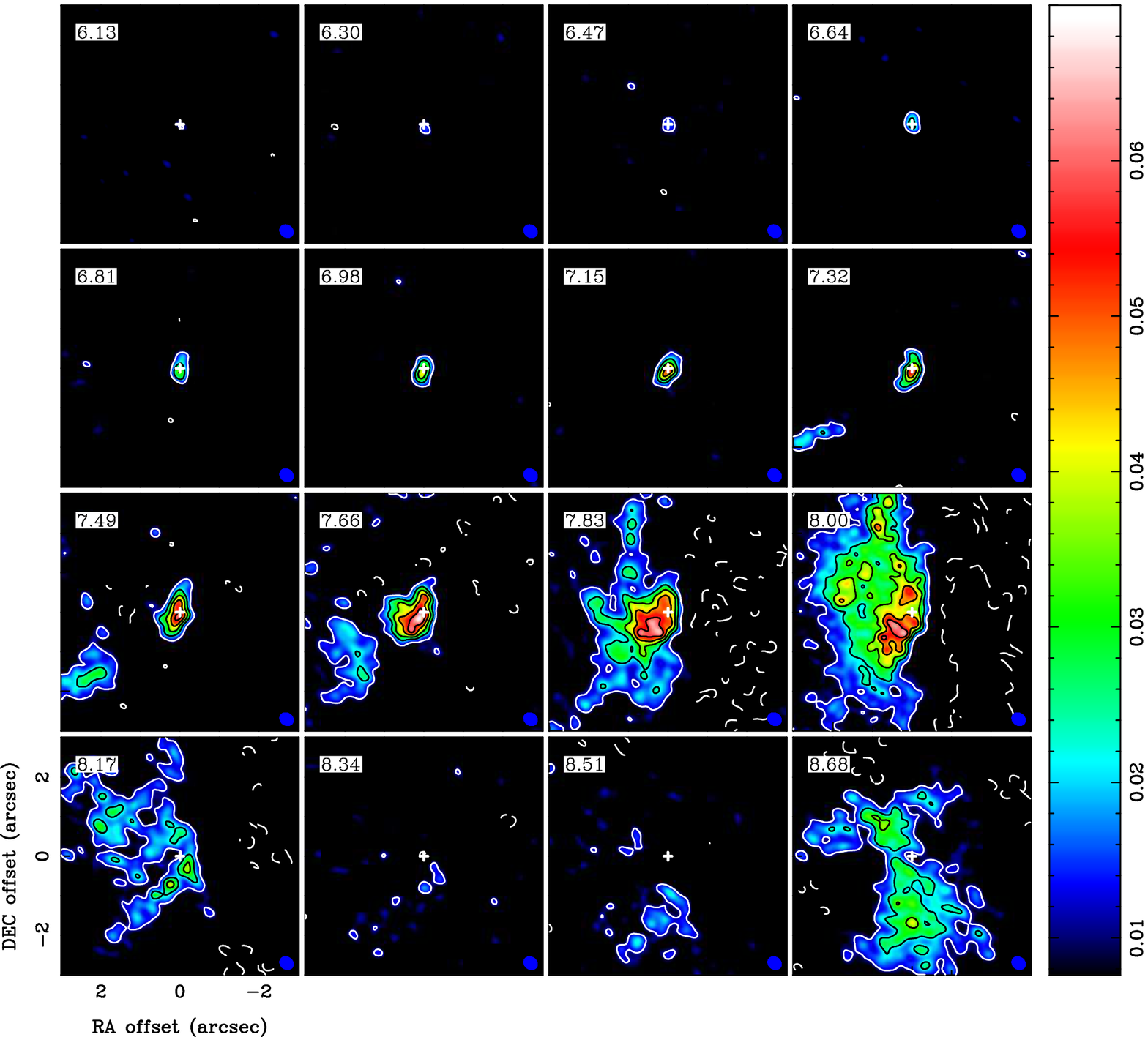}
\caption{Velocity channel maps of the C$^{18}$O (2--1) emission in B335. Crosses show the protostellar position. Filled ellipses at the bottom-right corners present the beam size. The central velocities of the channels are shown at the top-left corner in units of km s$^{-1}$. Contour levels are from 3$\sigma$, 10$\sigma$, 15$\sigma$, 20$\sigma$, and then in steps of 10$\sigma$, where 1$\sigma$ is 2.7 mJy beam$^{-1}$.}
\label{c18ochan1}
\end{figure}

\begin{figure}
\epsscale{1}
\plotone{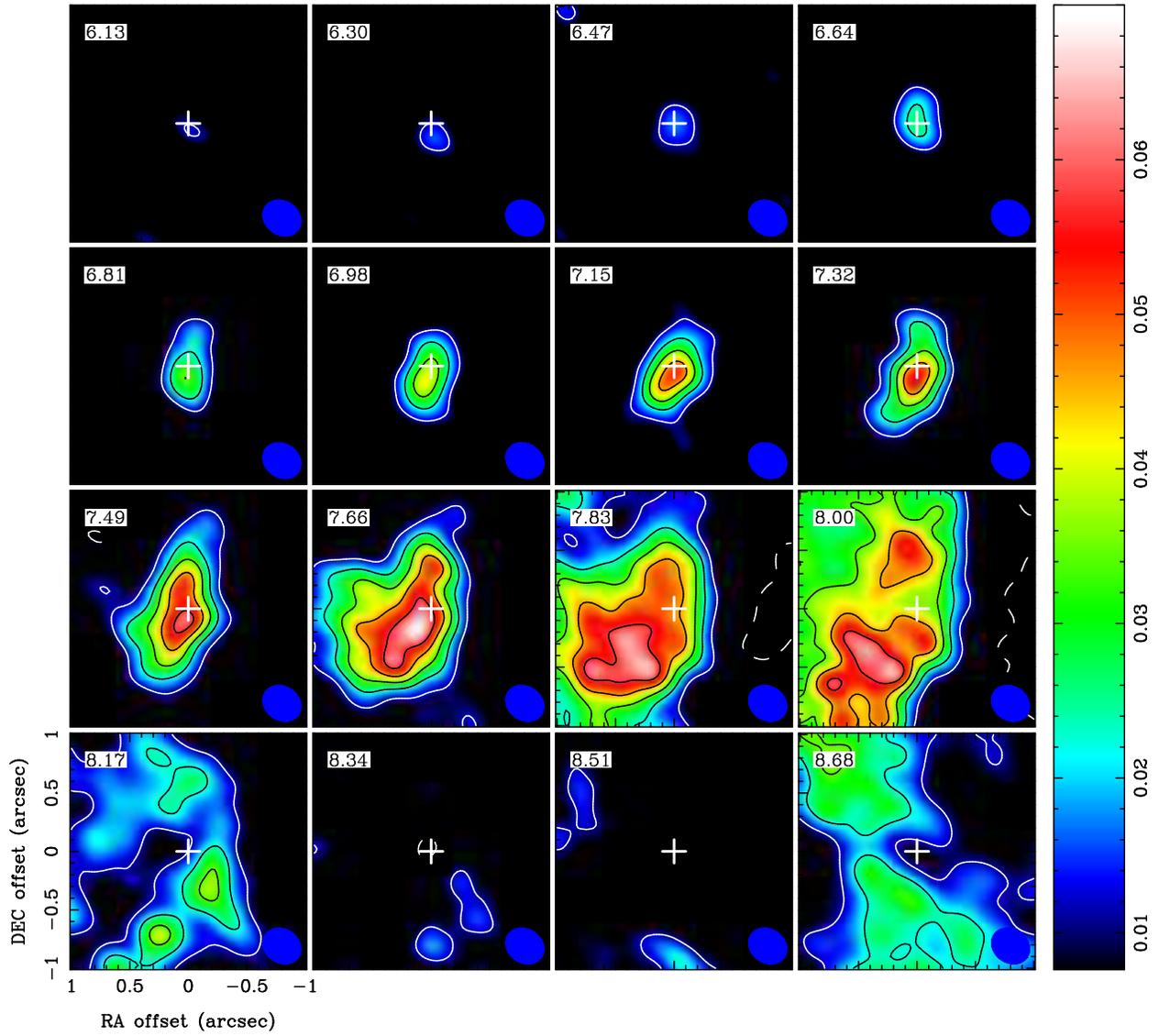}
\caption{Same as Figure \ref{c18ochan1} but zooming in on the central 2$\arcsec$ region.} 
\label{c18ochan1s}
\end{figure}

\begin{figure}
\epsscale{1}
\plotone{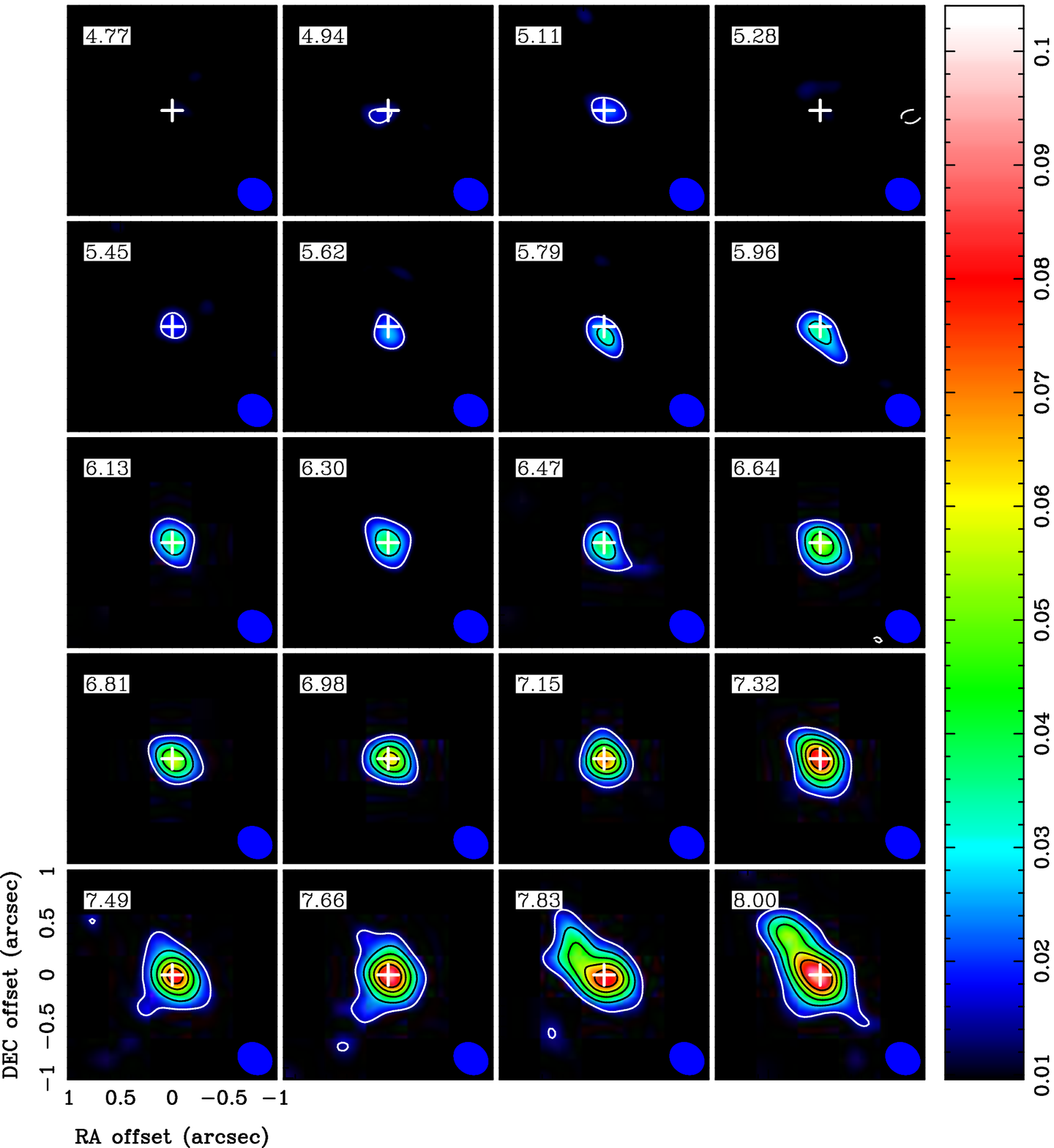}
\caption{Velocity channel maps of the SO (5$_6$--4$_5$) emission in B335. Crosses show the protostellar position. Filled ellipses at the bottom-right corners present the beam size. The central velocities of the channels are shown at the top-left corner in units of km s$^{-1}$. Contour levels are from 3$\sigma$, 10$\sigma$, 15$\sigma$, 20$\sigma$, and then in steps of 10$\sigma$, where 1$\sigma$ is 2.8 mJy beam$^{-1}$.}
\label{sochan1}
\end{figure}

Figure \ref{c18ochan1}, \ref{c18ochan1s}, and \ref{sochan1} present the velocity channel maps of the C$^{18}$O and SO emission, respectively.

\begin{figure}
\epsscale{0.8}
\plotone{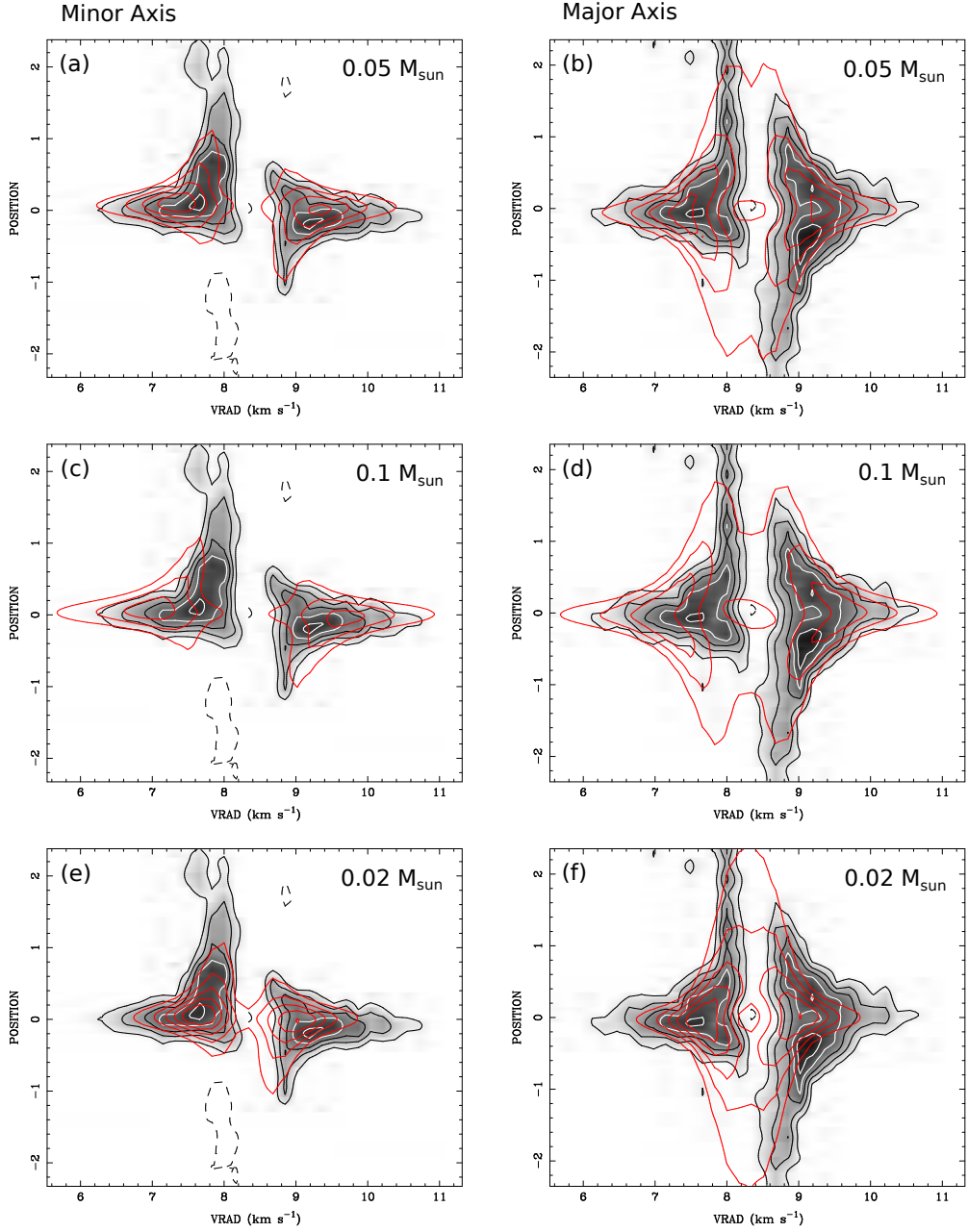}
\caption{Comparison between observed P--V diagrams of the C$^{18}$O emission along the minor (left panels) and major (right panels) axes of the 1.3 mm continuum emission (same as Figure \ref{c18opv}), and those computed from our kinematic models (red contours). 
Panels from top to bottom are for model protostellar masses of 0.05 $M_\sun$, 0.1 $M_\sun$ and 0.02 $M_\sun$. The vertical axis presents position in arcseconds. The horizontal axis is velocity in units of km s$^{-1}$. Contour levels are from 3$\sigma$ in steps of 3$\sigma$, where 1$\sigma$ is 3.8 mJy beam$^{-1}$.}
\label{modelpv}
\end{figure}

\begin{figure}
\epsscale{1}
\plotone{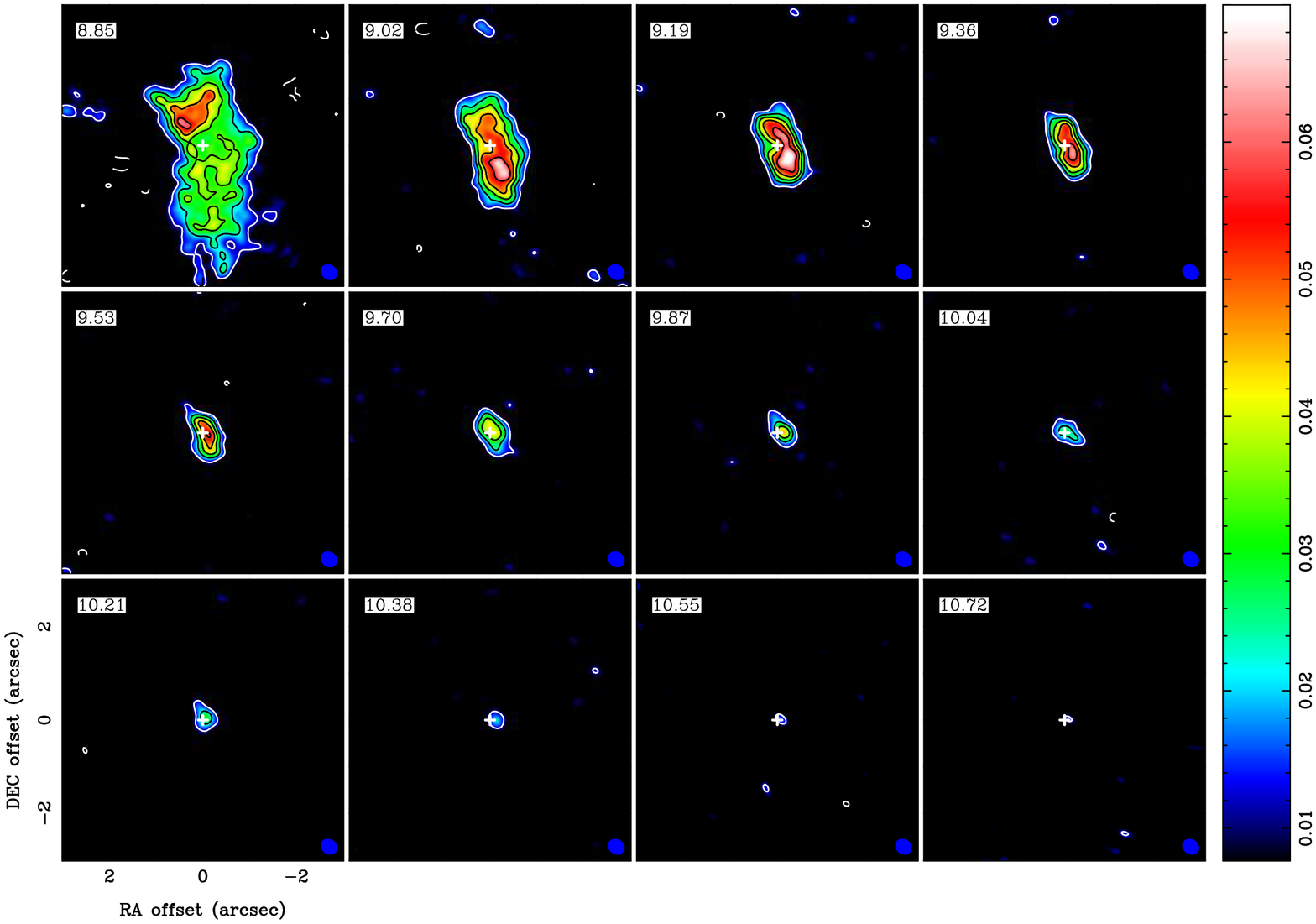}
Figure \ref{c18ochan1} -- Continued. 
\end{figure}

\begin{figure}
\epsscale{1}
\plotone{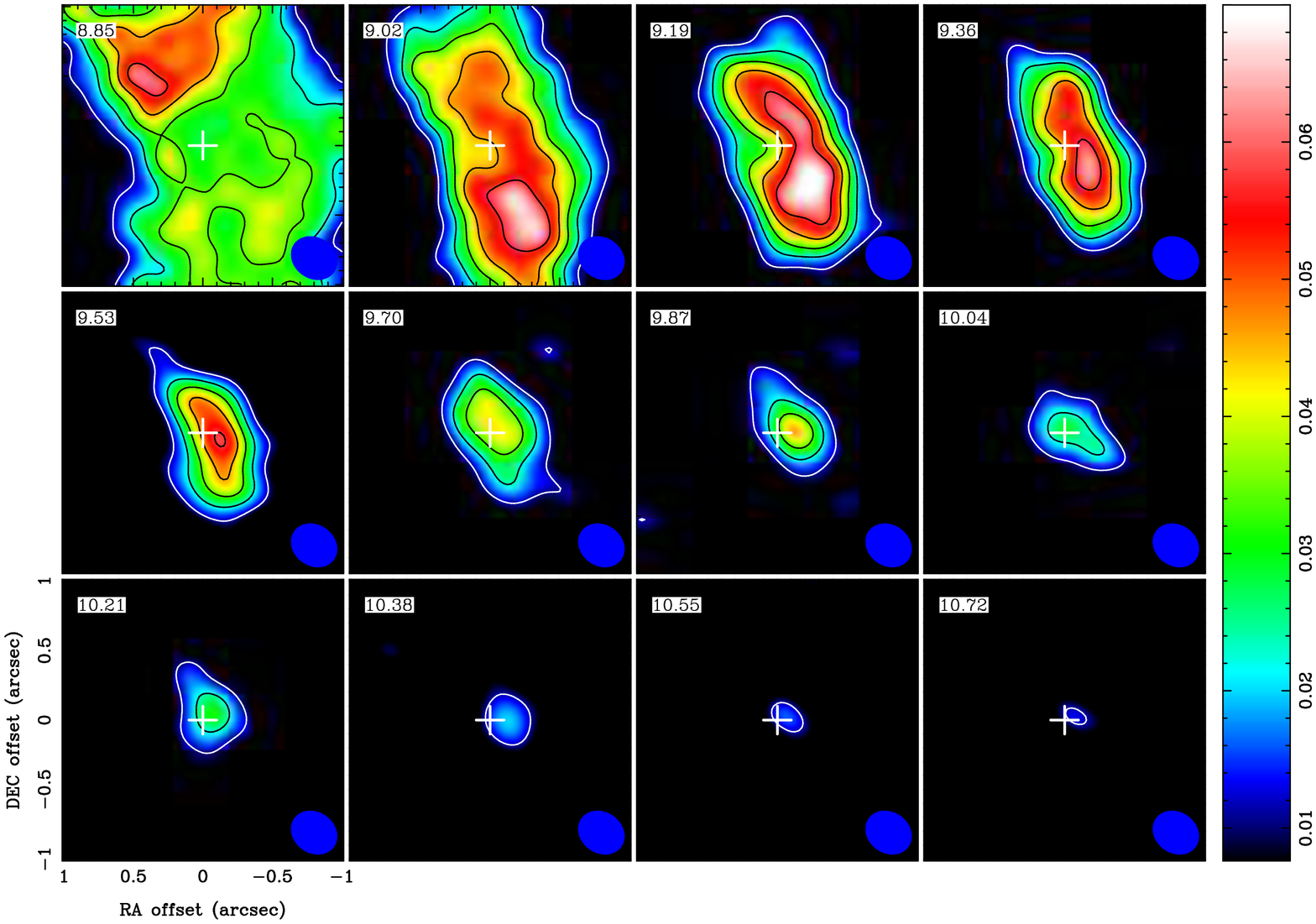}
Figure \ref{c18ochan1s} -- Continued. 
\end{figure}

\begin{figure}
\epsscale{1}
\plotone{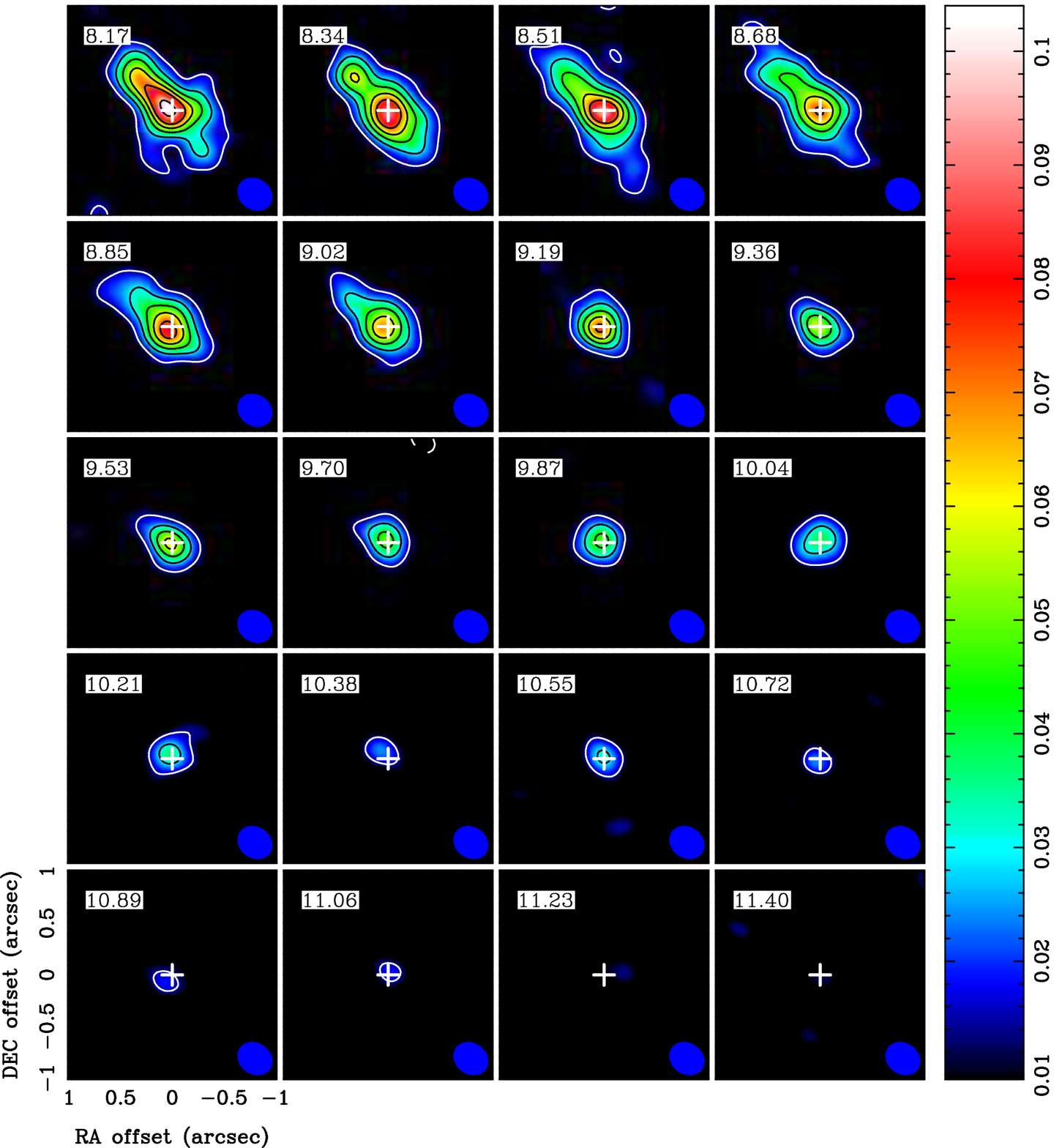}
Figure \ref{sochan1} -- Continued. 
\end{figure}

\clearpage

\begin{deluxetable}{lllcc}
\tablewidth{0pt}
\tablecaption{Summary of Imaging Parameters}
\tablehead{ & Weighting & Synthesized Beam & Noise Level & Velocity Resolution \\
& & & (mJy Beam$^{-1}$)  & (km s$^{-1}$) }
\startdata
Continuum & Natural & 0\farcs37 $\times$ 0\farcs32 (56\degr) & 0.24 & \\
 & Uniform &  0\farcs25 $\times$ 0\farcs22 (54\degr) & 0.26 & \\
 & $>$500 k$\lambda$ & 0\farcs19 $\times$ 0\farcs15 (50\degr) & 0.41 & \\
\hline
C$^{18}$O (2--1) & Robust ($r$ = 0.5) & 0\farcs34 $\times$ 0\farcs28 (55\degr) & 3.8\tablenotemark{a} & 0.17 \\
SO (5$_6$--4$_5$) & Robust ($r$ = 0.5) & 0\farcs34 $\times$ 0\farcs28 (54\degr) & 4.8\tablenotemark{a} & 0.17 \\
\enddata
\label{ob}
\tablenotetext{a}{Noise level per channel.}
\end{deluxetable}

\begin{deluxetable}{ccccc}
\tablewidth{0pt}
\tablecaption{1.3 mm Continuum Results}
\tablehead{ Component & Deconvolved Size (P.A.) & Total Flux & Mass  & $T_{\rm dust}$$\tablenotemark{a}$\\
& & (mJy) & ($M_\sun$) & (K)}
\startdata
Extended & $3\farcs34$ $\times$ $1\farcs41$ (25$\arcdeg$)  & 62.2 & 1.2 $\times$ 10$^{-2}$ & 30 \\
Compact & $0\farcs24$ $\times$ $0\farcs13$ (16$\arcdeg$) & 25.3 & (1.3--5.4) $\times$ 10$^{-3}$ & 30--100\\
Point Source & \nodata & 13.2 & (0.7--2.6) $\times$ 10$^{-3}$ & 30--100\\
\enddata
\tablenotetext{a}{Adopted dust temperature to estimate mass.}
\label{dust}
\end{deluxetable}

\end{document}